\documentclass[11pt]{article}

\usepackage{abstract}
\usepackage[utf8]{inputenc}
\usepackage[T1]{fontenc}
\usepackage{graphicx}
\usepackage{amsmath, amssymb, bm}
\usepackage{booktabs}
\usepackage{hyperref}
\usepackage{fancyhdr}
\usepackage[font={small,it}]{caption}	
\usepackage{subcaption}
\usepackage[export]{adjustbox}
\usepackage{booktabs}
\usepackage{makecell}
\usepackage{natbib}
\usepackage{abstract}
\usepackage{indentfirst}
\usepackage{comment}
\usepackage{xcolor}

\graphicspath{{img-stats/}}

\usepackage[a4paper, margin=1in]{geometry}

\title{Community-level core-periphery structures in collaboration networks}

\author{Sara Geremia\textsuperscript{1}, Domenico De Stefano\textsuperscript{1}, Michael Fop\textsuperscript{2} \\[1ex]
\small \textsuperscript{1}Department of Political and Social Sciences, University of Trieste, Italy \\
\small \textsuperscript{2}School of Mathematics and Statistics, University College Dublin, Ireland
}
\date{}

\begin{document}

\maketitle

\begin{abstract}
\noindent Uncovering structural patterns in collaboration networks is key for understanding how knowledge flows and innovation emerges. These networks often exhibit a rich interplay of meso-scale structures, such as communities, core-periphery organization, and influential hubs, which shape the complexity of scientific collaboration. The coexistence of such structures challenges traditional approaches, which typically isolate specific network patterns at the node level. We introduce a novel framework for detecting core–periphery structures at the community level. Given a reference grouping of the nodes, the method optimizes an objective function that assigns core or peripheral roles to communities by accounting for the density and strength of their inter-community connections. The node-level partition may correspond to either inferred communities or to a node-attribute classification, such as discipline or location, enabling direct interpretation of how different social or organizational groups occupy central positions in the network. The method is motivated by an application to a co-authorship network of Italian academics in three different disciplines, where it reveals a hierarchical core-periphery structure associated with institutional role, regional location, and research topics.
\end{abstract}


\clearpage

\section{Introduction}

The increasing collaboration and interdependence among researchers reflects the need for multidisciplinary approaches to address social, political, economic, and technological challenges \citep{Wuchty2007TheKnowledge}. Uncovering structures in collaborative research networks is crucial for understanding how ideas spread and capabilities evolve, ultimately driving scientific productivity and the generation of new knowledge \citep{DeStefano2011IssuesNetworks}. 
To deepen understanding of these complex mechanisms, researchers have developed statistical approaches aimed at identifying structures in collaboration networks, with large-scale co-authorship network analyses enabled by the increasing availability of bibliometric data.\citep{Donthu2021HowGuidelines}. 

A growing body of work focuses on meso-scale network structures, with community detection being a leading theme. Communities are generally understood as groups of nodes with similar connectivity patterns or shared characteristics, although the lack of a universal definition has led to the development of numerous community detection approaches, either heuristic or model-based (see e.g. \cite{Fortunato2016CommunityGuide, Lee2019AClustering} for recent surveys). The first category includes widely used methods such as modularity-based algorithms like Louvain \citep{Blondel2008FastNetworks} or flow-based methods like Infomap \citep{Rosvall2007AnNetworks}. Model-based approaches partition nodes into distinct blocks, and the probability of an edge connecting any two nodes is determined solely by their block memberships. The most popular probabilistic model used to generate and represent network structures is the stochastic block model \citep[SBM,][]{Holland1983StochasticSteps, Wasserman1987StochasticAssessment}. 

In research and innovation systems, communities are frequently shaped around influential actors, but also thematic areas or regional collaborations, forming social circles \cite{Alba1978ELITECIRCLES, Alba1976}. Acknowledging the interdependencies among community members is crucial for understanding how collaboration evolves within the community.
Equally important is to examine how community structures interact with other organizational patterns, such as disassortativity, core-periphery dynamics, or the presence of hubs \citep{Fortunato2010CommunityGraphs, Legramanti2022ExtendedNetworks, Kojaku2018Core-peripheryNetwork}. 

While community structures in collaboration networks have been extensively studied \citep{Newman2004FindingNetworks, Luzar2014CommunityNetwork, Menardi2022Density-basedNetworks}, relatively less attention has been paid to their arrangement in core-periphery structures \citep{Zelnio2012IdentifyingScience, Karlovcec2016Core-peripherySlovenia, Sedita2020TheLiterature, Wedell2022CenterperipheryCommunities}. This structure is used to describe systems where a cohesive, central group, the {\em core}, interacts densely within itself, while a more loosely connected group, the {\em periphery}, maintains ties mainly with the core \citep{Yanchenko2023Core-peripheryExposition, Tang2019RecentSurvey}. 

 Despite the potential interplay between community and core-periphery structures, their combined presence in social networks has rarely been explored \citep{Legramanti2022ExtendedNetworks}.
An initial strategy to address this gap is to consider methods capable of capturing more general block structures in network data. Notably, SBMs provide flexible frameworks for uncovering core–periphery and hierarchical structures \citep{Gallagher2021ANetworks, Yanchenko2023Core-peripheryExposition}.  

Building on this idea, we develop a core-periphery detection framework specifically designed to detect nested meso-scale structures. Unlike traditional methods that identify core-periphery organization at the node level, we extend the method to identify core and peripheral groups of nodes. Specifically, assuming a community structure in the network, we focus on a binary partition of the communities, inspired by the two-block model of \cite{Borgatti2000ModelsStructures}. To the best of our knowledge, this is the first approach to explicitly identify nested core-periphery organization in collaboration networks, revealing how such structures emerge between communities rather than within them. In doing so, we introduce a novel community-level framework that redefines the focus of core–periphery detection beyond individual nodes.

In Section~\ref{sec:example} we present the motivating collaboration network that inspired our study, along with the associated research questions. Section~\ref{sec:background} provides the background: we review the related literature (Section~\ref{sec:literature}) and introduce the new community-level core-periphery network structure under investigation (Section~\ref{sec:structure}). In Section~\ref{sec:method}, we formally define the community-based core-periphery detection problem and describe our proposed approach. Simulation studies in Section~\ref{sec:simulation} assess the performance of the proposed approach. Section~\ref{sec:application} discusses the findings of the collaboration network analysis. Finally, Section~\ref{sec:discussion} concludes with a discussion and directions for future research.

\subsection{Co-authorship network of Italian academic scholars}\label{sec:example}

Our approach is motivated by a case study on co-authorship within the scientific community of Italian academic scholars, as recorded in the Italian Ministry of University (MUR) roster in December 2022. The data concern collaborations in a co-authorship network involving statisticians over 10 years, from 2012 to 2022.

The edges in the co-authorship network represent relationships derived from the shared collaborations in the paper-author bipartite network. Hence, an edge between two authors is formed if both are involved in at least one common scientific paper. The edges can carry weights that reflect the number of shared papers, and edge weights indicate the number of shared publications. In this setting, the weight distribution is highly skewed, with a few author pairs accounting for a disproportionate number of collaborations. Moreover, the weights are largely influenced by factors such as project size and journal authorship rules, which may capture institutional or disciplinary proximity rather than relational strength. For these reasons, we base our analysis on the unweighted version of the network. 

Although the academic statistics community collaborates across multiple disciplines, we focus on relationships between statistics, sociology, and business to capture the strongest collaborations in applied quantitative research within the social and economic domains. As shown by \citep{Stefano2023CollaborationStatisticians,fabbrucci2025unveiling}, Economic Statistics (S/03) and Social Statistics (S/05) are particularly integrated with economics, finance, and social sciences. 
Moreover, in many cases, statistics is institutionally located within economics or social sciences departments, reinforcing the relevance of this focus.
This disciplinary focus is enabled by the availability of detailed information on academic subfields through the Italian {\em Settore Scientifico Disciplinare} (SSD, i.e. Scientific Discipline Sector) classification system. 

Additional information for our analysis includes author and publication data sourced from Scopus and MUR. In particular, we take into account authors’ location (as determined by their affiliation), academic role, and the thematic focus of their publications. We are interested in assessing whether geographical and cultural proximity, shared research interests, as well as academic seniority, drive collaboration \citep{Katz1994GeographicalCollaboration}. 

Table~\ref{tab:tab1} presents the main characteristics of the co-authorship network. The dataset comprises 2,691 academic scholars collaborating on 1,809 papers, with a network density of 0.09\%, reflecting sparse collaboration across the network. 
Statisticians represent 30.7\% of the network, and exhibit the highest connectivity and bridging role, with an average degree and betweenness of 5.06 and 5883.28, respectively, and a density of 0.20\%, suggesting strong group interconnections. In contrast, researchers from business and sociology are more weakly connected, with edge densities of 0.05\% and 0.04\%, respectively.

\begin{table}[b!]
    \centering
\begin{tabular}{llrrrrrr}
  \toprule
   &  &   & N & \% & \makecell{Average\\Degree} & \makecell{Average\\Betweenness} & \makecell{Density (\%)}\\    
      \cmidrule{1-8}
    & Total & & 2691 & & 2.34 & 2917.98 & 0.09 \\
     \cmidrule{1-8} 
 Field & Statistics &  & 827 & 30.73 & 5.06 & 5883.28 & 0.20 \\ 
 & Business &  & 821 & 30.51 & 1.26 & 1489.36 & 0.05\\ 
   & Sociology &  & 1043 & 38.76 & 1.02 & 1691.31 & 0.04\\
    \cmidrule{1-8}     
 Role  & Full Professor &  & 748 & 27.80 & 3.20 & 4662.07 & 0.12\\ 
 & Associate &  & 1134 & 42.14 & 2.25 & 2696.91 & 0.08\\ 
   & Researcher &  & 809 & 30.06 & 1.67 & 1615.27 & 0.06 \\ 
     \cmidrule{1-8} 
Location 
   & North-West &  & 687 & 25.53 & 2.23 & 2546.70 & 0.08\\ 
   & North-East &  & 591 & 21.96 & 2.29 & 3251.66 & 0.09\\ 
   & Center &  & 689 & 25.60 & 2.39 & 3369.73 & 0.09\\    
   & South &  & 517 & 19.21 & 2.49 & 2618.23 & 0.09\\ 
   & Islands &  & 207 & 7.69 & 2.28 & 2442.49 & 0.09\\ 
   \bottomrule
\end{tabular}
    \caption{Co-authorship network characteristics}
    \label{tab:tab1}
\end{table}

As expected, collaboration patterns vary by academic seniority. Full Professors show the highest centrality, playing a central role in facilitating connections. 
Collaboration levels appear relatively consistent across regions, with minor variations.
Researchers from the Center and North-East regions report the highest average betweenness, while those from the Islands show the lowest connectivity. 

With a transitivity value of 0.26, the network demonstrates a moderate tendency toward community formation. At the same time, the network exhibits a heterogeneous degree distribution, with highly connected nodes that may belong to different communities and create hierarchical patterns. 
These patterns suggest the coexistence of different meso-scale structures: strong intra-disciplinary ties define cohesive subnetworks, whereas bridging roles of senior academics, statisticians, and certain regions suggest a core–periphery dynamic. Overall, thematic, institutional, and geographical features combine to form a layered architecture in which communities, cores, and hubs jointly shape collaboration and knowledge flow.

Since co-authorship networks inherently reflect the interdependence of groups of collaborators, we ask whether these communities are equally integrated into the broader network or whether some remain relatively isolated, with members collaborating primarily within their own group. If such isolation exists, our goal is to characterize these communities and investigate whether their decentralization is shaped by factors such as members’ institutional location, academic role, or research focus. Our main contribution lies in formalizing the problem and showing that it can be addressed as a core–periphery partition of the identified communities. In addition, we develop a framework capable of detecting this community-level core–periphery organization.

\section{Core-periphery structures in networks}\label{sec:background}

\subsection{Relevant literature}\label{sec:literature}

The concept of {\em core-periphery} has its roots in economic and sociopolitical theories \citep[e.g.][]{Prebisch1949TheProblems, Galtung1971AImperialism} and was formalized in network science through blockmodeling approaches in the 1970s \citep{Breiger1975AnScaling, Mullins1977TheStudy, White1976SocialPositions}. Since then, it has become a central idea for describing heterogeneity in networks, with applications spanning the social, economic, and biological sciences \citep{Hidalgo2007TheNations,Rombach2012Core-PeripheryNetworks,Bassett2013Task-BasedDynamics}. However, as recent surveys emphasize \citep{Tang2019RecentSurvey, Yanchenko2023Core-peripheryExposition}, there is no single, universally accepted definition of what constitutes a core or a periphery. Instead, multiple conceptualizations coexist, each relying on different assumptions about how core and peripheral nodes should relate to one another. As a consequence, researchers are often faced with methods that -- while all labeled ``core–periphery detection'' -- encode distinct and sometimes incompatible visions of the concept. 

Two main definitions, each derived from exemplary models, dominate the literature.
The first is associated with the layered (or k-core) model, which interprets coreness as a gradual property rather than a strict dichotomy. In this framework, approaches for core-periphery detection embed nodes in nested layers, with inner layers representing the most central core \citep{Batagelj2003AnNetworks, Goltsev2006KEffects, Hebert-Dufresne2016Multi-scaleDecomposition, Gallagher2021ANetworks}. These approaches, also referred to as transport-based methods, exploit flow dynamics or path structures, typically relying on geodesic distances or random walks to assign coreness scores \citep{Lee2014Density-basedNetworks, Cucuringu2016DetectionPaths}. They allow for multiple levels of peripheral integration and have been widely applied to capture hierarchical organization in large-scale networks.
 
The second major paradigm corresponds to the two-block (hub-and-spoke) model, formalized by \cite{Borgatti2000ModelsStructures}. In this formulation, core nodes are densely connected both among themselves and with peripheral nodes, whereas peripheral nodes maintain only sparse interconnections. This hub-and-spoke view assumes overall network connectivity, with the periphery integrated through its ties to the core. The \cite{Borgatti2000ModelsStructures} framework remains one of the most widely adopted approaches to core-periphery detection in practice, despite subsequent developments and alternative formulations \citep{Boyd2006ComputingData, Brusco2011AnProblem, Lip2011AProblem, Zhang2015IdentificationNetworks,  Gallagher2021ANetworks, Estevez2025RevisingCores}. A key strength of this family of models is its simplicity: nodes are dichotomized into either core or periphery, enabling straightforward interpretation. Moreover, the two-block model has close connections with the SBM\citep{Karrer2011StochasticNetworks}, as both evaluate connectivity patterns against an idealized density-based structure. 

Despite its popularity, the \cite{Borgatti2000ModelsStructures} framework presents several limitations when applied to real-world networks. \cite{Kojaku2018Core-peripheryNetwork} argued that simple hub-and-spoke structures lack the flexibility to account for interesting patterns other than those already explained by degree heterogeneity. 
Instead, they suggest that identifying a meaningful core-periphery organization requires identifying additional substructures within the same network, such as communities, bipartite patterns, or overlapping core–periphery pairs. Their work specifically focuses on uncovering the latter \citep{Kojaku2017FindingNetworks, Kojaku2018Core-peripheryNetwork}.
Other studies have examined similar multi-core structures, where several dense, mutually interacting cores coexist alongside peripheral regions \cite{Rombach2012Core-PeripheryNetworks, Zhang2015IdentificationNetworks}. Such a multi-core structure arises naturally in SBM formulations, particularly in their hierarchical and microcanonical variants \citep{oto2014HierarchicalModels,oto2019BayesianBlockmodeling,Come2021HierarchicalLikelihood}. By linking community detection with core–periphery decomposition, these models show that communities themselves can act as cores or peripheries relative to one another. 

A structure closely related to the multi-core core-periphery organization is the rich-club or elite-circle structure \citep{Alba1978ELITECIRCLES, Zhou2004TheTopology, Colizza2006DetectingNetworks}, which describes the tendency of high-degree nodes to form densely interconnected groups. In social networks, this phenomenon often signals the emergence of an ``oligarchy'' of influential actors who dominate communication, in contrast to decentralized structures formed by loosely connected communities. Unlike the simple degree assortativity of the two-block model, the rich-club effect cannot be explained by degree heterogeneity alone and can arise in both assortative and disassortative networks \citep{Newman2002AssortativeNetworks}. This distinction links rich-club theory with multi-core structures, as both point to overlapping or interacting dense subgroups within a larger network. 

Despite this rich literature, the interplay between community structure and core–periphery organization has received limited systematic treatment. While some studies have hinted at their coexistence, to the best of our knowledge, no existing framework explicitly identifies core and peripheral communities. 
To address this gap, we introduce a novel approach that generalizes the traditional two-block, node-level core–periphery model to the community level, enabling the detection of core and peripheral organization among communities.

\subsection{Community-level core-periphery structures}\label{sec:structure}

\begin{figure}[t!]
    \centering
    
    \begin{subfigure}[t]{0.3\textwidth}
        \centering
        \includegraphics[width=\textwidth]{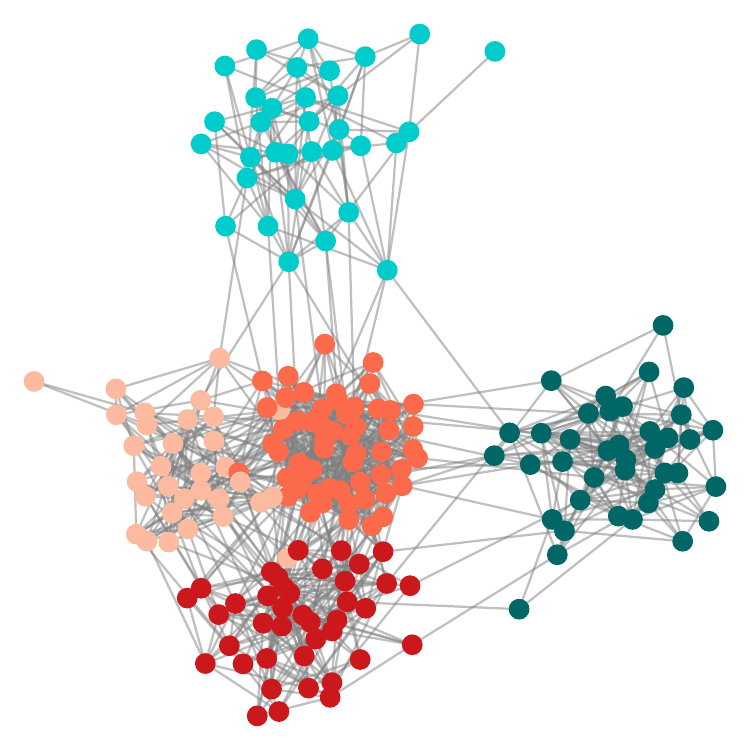}
        \caption{}
        \label{fig:sub1}
    \end{subfigure}
    \hfill
    \begin{subfigure}[t]{0.3\textwidth}
        \centering
        \includegraphics[width=\textwidth]{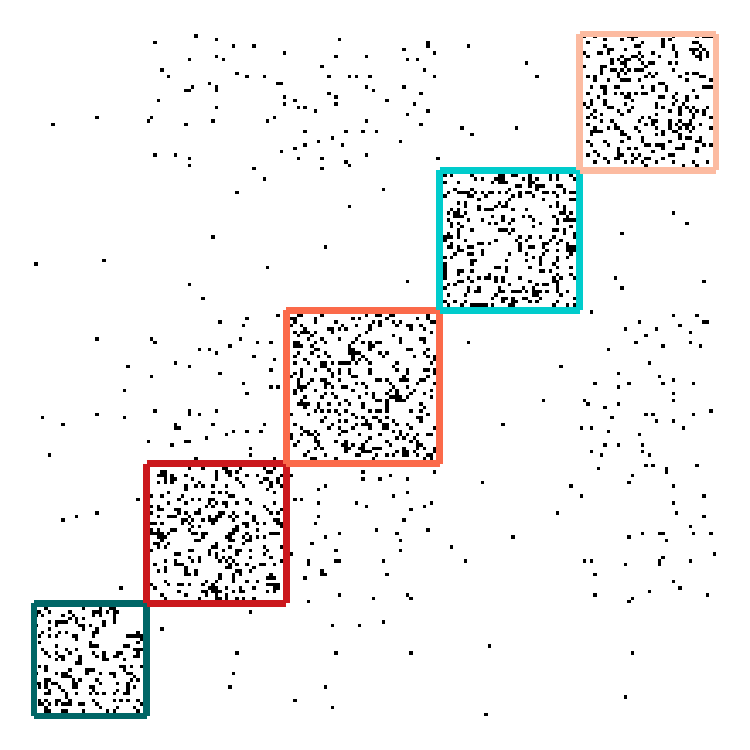}
        \caption{}
        \label{fig:sub2}
    \end{subfigure}
    \hfill
    \begin{subfigure}[t]{0.3\textwidth}
        \centering
        \includegraphics[width=\textwidth]{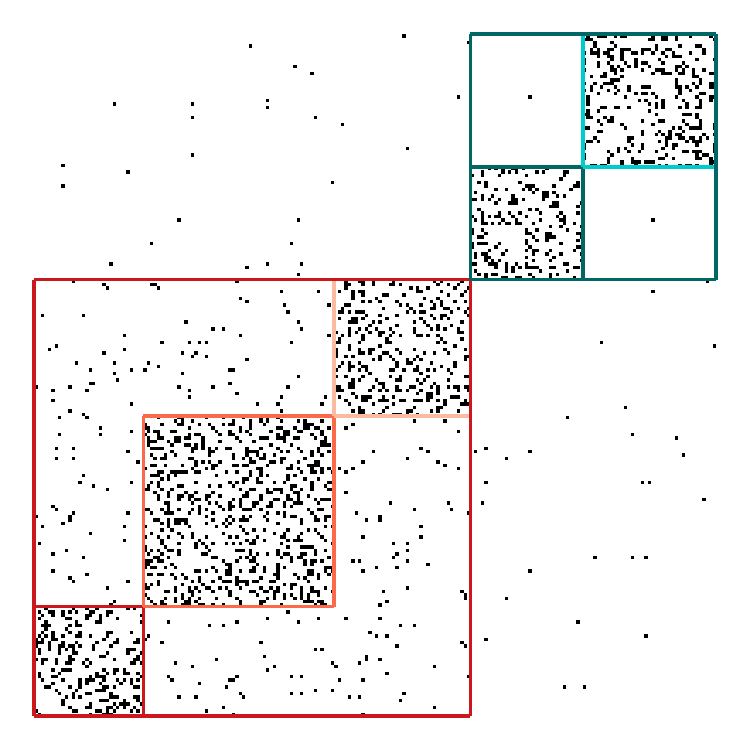}
        \caption{}
        \label{fig:sub3}
    \end{subfigure}

    \caption{(a) Example of a network partition in 5 communities with core-periphery roles;  (b) Adjacency matrix representing a community partition of the nodes; (c) Adjacency matrix representing a community-level core-periphery partition. Blue color denotes periphery, while red color denotes core.}
    \label{fig:fig1}
\end{figure}

While most analyses focus on the roles of individual nodes, our approach shifts attention to how entire communities of interdependent actors are positioned within broader network architectures. This perspective highlights two nested levels of network structure:

\begin{itemize}
    \item \textbf{Community partition}: Nodes are organized into densely connected groups (communities) that exhibit relatively sparse connectivity between them (Figure~\ref{fig:sub2}).
    \item \textbf{Community-level core–periphery partition}: Communities are hierarchically organized, distinguishing those that form the core, which are densely interconnected and central, from those in the periphery, which are sparsely connected and with limited links to other peripheral groups (Figure~\ref{fig:sub3}). 
\end{itemize}

The proposed community-level core-periphery detection approach aims to identify groups of nodes based on their relationships and the hierarchical structure among those groups. 
In many complex social systems, communities can be understood as circles of interdependent actors, and recognizing these interdependencies is crucial for: explaining how information, resources, or influence circulate; how collective dynamics are sustained; and how certain groups come to occupy more central positions within the broader network. In this framework, central positions are typically held by communities that are densely interconnected and act as bridges across different parts of the system (i.e., core communities). Peripheral communities, by contrast, are weakly connected among them, with their participation in the system often mediated by their relations with the core.

Shifting the attention from individual nodes to groups of nodes enables the identification of collective behavior and shared structural roles. This perspective reflects the fact that many outcomes -- such as innovation, diffusion, or resilience -- are driven by interactions among groups rather than isolated actors. By treating communities as the analytical unit, we capture meso-scale dynamics that more accurately represent how functional subsystems operate within a network.

The proposed community-level core-periphery detection approach substantially enhances interpretability: by grouping nodes based on shared structural or attribute-based characteristics, it becomes possible to directly relate network positions to real-world social or organizational entities. For example, categorical attributes, such as geographical location or functional role, can naturally serve as cluster labels, allowing us to examine how different types of groups align with core or peripheral positions. This helps identify which subsystems or domains serve as key hubs within the overall structure.

Moreover, by analyzing inter-community connection patterns, we can assign and interpret intra-community leadership. This means that the density or strength of inter-community connections helps identify which communities, and which nodes within those communities, act as central influencers.  

An additional motivation for identifying core communities stems from empirical observations in SBMs, which, in networks with substantial degree heterogeneity, often reveal clusters with hub-like behavior, that is, groups that maintain dense connections across the network \cite{Karrer2011StochasticNetworks}. Rather than interpreting these highly connected clusters in isolation, grouping them into a unified core block provides a clearer understanding of their structural role.

Finally, adopting a community-level perspective improves both scalability and robustness in large, complex networks. Aggregating nodes into communities reduces noise from individual-level variability—such as the disproportionate influence of highly connected hubs—and facilitates the detection of persistent structural patterns across scales. This is particularly valuable in networks that span multiple domains, organizations, or regions, where individual-level connectivity may obscure broader systemic dynamics.

\section{Community-level core-periphery detection}\label{sec:method}

In this section, we present the proposed community-level core-periphery detection framework. The framework is based on an objective function specifically designed to identify community-level core–periphery structures under the constraint that the set of communities is partitioned into exactly one core group and exactly one peripheral group, each containing at least two communities. 
This formulation is inspired by the objective function introduced in \cite{Cucuringu2016DetectionPaths}, but extends it by replacing node-level coreness with measures that summarize connectivity within and between communities.

The procedure consists of three main steps.
\begin{description}
    \item[Step 1.] Partition the node set of the network into $K$ communities. These communities may be the result of the application of a community detection method appropriate for the data and the specific research task. Alternatively, the communities can correspond to an existing classification of the nodes, for instance, one corresponding to a particular node-level categorical attribute.
    
    Let $Q = \{q_1, \ldots, q_k, \ldots q_K\}$ denote the set of communities.

    \item[Step 2.] Estimate the connectivity between the each pair of communities $q_k$ and $q_l$: 
    $$\hat{\theta}_{kl} = \frac{m_{kl}}{n_k n_l}, \quad k, l = 1 ,\ldots, K, ~~k \neq l,$$
    where $m_{kl}$ is the number of observed edges between the two communities, and $n_{k}$, $n_{l}$ are their sizes (i.e. number of nodes assigned to each community). Note that this estimator is consistent with the connectivity parameter estimates of a standard block model or SBM for binary networks \citep{White1976SocialPositions,  Holland1983StochasticSteps,Karrer2011StochasticNetworks}.

    \item[Step 3.] Evaluate the quality of candidate core–periphery partitions of the communities. This is achieved by computing the value of an objective function, $\phi(\mathbf{z};\hat{\bm{\Theta}})$, where $\hat{\bm{\Theta}}$ denotes the estimated connectivity matrix with off-diagonal entries $\hat{\theta}_{kl}$ and null diagonal, and $\mathbf{z}$ is an indicator vector specifying whether a each community belongs to the core or the periphery. The function $\phi(\mathbf{z}; \hat{\bm{\Theta}})$ is then optimized with respect to $\mathbf{z}$ over all admissible partitions of the communities into core and periphery groups.
\end{description}

We now describe Step 3 in more detail. Formally, each community $q_k$ is associated with a binary variable $z_k \in \{0, 1\}$, where $z_k = 1$ if community $q_k$ belongs to the core, while $z_k = 0$ if it belongs to the periphery.

We impose the constraints:
$$
\sum_{k=1}^K z_k \geq 2 \quad \text{and} \quad \sum_{k=1}^K (1 - z_k) \geq 2,
$$
ensuring that both the core and the periphery contain at least two communities. This excludes degenerate cases with core or periphery comprising only one community. The partitioning of $Q$ into core and peripheral communities is thus determined by the vector $\mathbf{z} = (z_1,\ldots,z_k,\ldots, z_K)$.

The connectivity strengths between pairs of communities, captured in $\hat{\bm{\Theta}} = [\hat{\theta}_{kl} ]$, reflect both the topology (presence or absence of edges) and the intensity (edge frequency) of inter-community interactions. To evaluate a given partition $\mathbf{z}$, we define the following sets of nonzero inter-community connectivity values extracted from $\hat{\bm{\Theta}}$, which encode connectivity among core and peripheral communities, respectively:
$${\cal T}_{c} = \left\{ \hat{\theta}_{kl} : k < l,\ z_k z_l = 1,\ \hat{\theta}_{kl} > 0 \right\},$$
$${\cal T}_{p} = \left\{ \hat{\theta}_{kl} : k < l,\ (1 - z_k)(1 - z_l) = 1,\ \hat{\theta}_{kl} > 0 \right\}.
$$
The objective function $\phi(\mathbf{z};\hat{\bm{\Theta}})$ is defined as
\begin{equation}\label{eq:objective}
    \phi(\mathbf{z};\hat{\bm{\Theta}}) = \big(\delta_c + \mu_c - \sigma_c\big) - \big(\delta_p + \mu_p + \sigma_p\big),
\end{equation}
where:
\begin{itemize}
    \item $\delta_c, \delta_p$ denote the densities of nonzero edges in the core and periphery, respectively:
    $$
    \delta_c = \frac{|{\cal T}_{c}|}{\sum_{k<l} z_k z_l}, \qquad
    \delta_p = \frac{|{\cal T}_{p}|}{\sum_{k<l} (1 - z_k)(1 - z_l)};
    $$

\item $\mu_c, \mu_p$ are mean edge frequencies in the core and periphery,
$$
\mu_c = \dfrac{1}{|{\cal T}_{c}|} \sum_{\hat{\theta}_{kl} \in {\cal T}_c}\hat{\theta}_{kl}, \qquad
\mu_p = \dfrac{1}{|{\cal T}_{p}|} \sum_{\hat{\theta}_{kl} \in {\cal T}_p}\hat{\theta}_{kl};
$$

\item $\sigma_c, \sigma_p$ are standard deviations (sd) of the edge frequencies,

$$
\sigma_c = \sqrt{ \dfrac{1}{|{\cal T}_{c}| - 1} \sum_{\hat{\theta}_{kl} \in {\cal T}_c} (\hat{\theta}_{kl} - \mu_c) }, \qquad
\sigma_p = \sqrt{ \dfrac{1}{|{\cal T}_{p}| - 1} \sum_{\hat{\theta}_{kl} \in {\cal T}_p} (\hat{\theta}_{kl} - \mu_p) }.
$$
\end{itemize}

Zero entries are excluded from ${\cal T}_c$ and ${\cal T}_p$ to ensure that the summary statistics capture only the strength and variability of existing inter-community connections, while the overall proportion of such connections is already reflected in the density terms. This choice is in line with the definition of periphery, where links among peripheral communities are expected to be sparse, and when present, as weak as possible. By considering only nonzero entries, the method is able to identify configurations in which peripheral links exist but remain weak, thereby emphasizing interaction intensity rather than sole presence or absence.

By construction, the function in Equation \eqref{eq:objective} increases when core communities are denser, more strongly connected, and more homogeneous than peripheral ones. 
A high value of $\delta_c$ and a low value of $\delta_p$ indicate a denser core and a sparser periphery, consistent with a typical core-periphery structure. We use the mean ($\mu$) as a measure of the central tendency of the edge frequency distribution. Maximizing $\mu_c$ while minimizing $\mu_p$ reinforces the contrast between the interaction intensity within the core and that within the periphery.
We use the standard deviation ($\sigma$) as a measure of dispersion. The $\sigma_c$ and $\sigma_p$ terms penalize high variability in edge frequencies, preventing solutions in which the core consists only of a few unusually strong links and the periphery of links of very heterogeneous strength. This also discourages the opposite case, with a small periphery with extremely weak links and a core with heterogeneous interconnections.

The groups are determined endogenously by the data, with no constraints on their size. We do not require balanced groupings, and the number of core and peripheral communities can vary as long as both sets include at least two elements. 

The optimization of the function in Equation \eqref{eq:objective} with respect to $\mathbf{z}$ is a combinatorial problem, as it requires evaluating discrete partitions of communities into core and periphery. The search space grows exponentially with the number of communities, making exhaustive enumeration infeasible even for moderately sized networks.
To address these challenges, we use a genetic algorithm \citep{Scrucca2013GA:R}, inspired by natural selection. Genetic algorithms are particularly well-suited to this problem because they can efficiently explore large and complex search spaces, balance exploitation of promising solutions with exploration of new ones, and avoid getting trapped in suboptimal local optima. This approach allows us to approximate the optimal community-level assignment between the center and the periphery, while maintaining computational feasibility and robustness across different network structures.

\section{Simulation studies}\label{sec:simulation}

To assess the empirical performance of the proposed approach, we conduct a simulation study based on varying networks generated from SBMs. As highlighted in the introduction, while the SBM is widely recognized for its ability to generate networks with community structures, it is also sufficiently flexible to accommodate more complex patterns, including core–periphery configurations.
In particular, the generative model developed for this study is explicitly designed to simulate networks where both community structure and inter-community heterogeneity coexist.

We aim to evaluate how effectively the proposed approach can recover the true partitioning of communities into core and periphery, both in terms of objective function behavior and core-periphery detection performance.

\subsection{Simulation design}

Networks are generated to reproduce realistic patterns of inter-community collaboration. Different network sizes, $n = \{100, 200, 500, 1000 \}$ and numbers of communities, $K = \{5, 10, 15 \}$ are considered.  A parameter $\lambda = \{25, 50, 75\}$ controls the percentage of communities belonging to the core. The remaining communities are treated as peripheral. Each node is then assigned to one of these communities according to a probability distribution, which is either uniform (equal-sized communities) or drawn from a Dirichlet distribution to reflect size heterogeneity across communities. 

Once both the community and community-level core-periphery membership are determined, we define connection probabilities within and between communities. Nodes belonging to the same community are connected with a high probability. Inter-community connections are formed on the basis of the core-periphery classification of the involved communities: core-to-core connections are most likely, core-to-periphery connections are less likely, and periphery-to-periphery connections are least likely. Further details are provided in the Appendix~\ref{sec:appendix:sim_details}.

To ensure realistic scenarios, we exclude certain parameter combinations that could produce degenerate structures: specifically, small networks (with 100 or 200 nodes) were not paired with a large number of communities ($K = 20$), and large networks (with 500 or 1000 nodes) were not paired with very few communities ($K = 5$). The collection of parameter combinations provides 96 distinct scenarios; for each scenario, we simulate 100 independent network data realizations. Data are generated, ensuring that every community includes at least two nodes. 

The performance of core-periphery detection at the node level is evaluated using standard metrics: balanced accuracy (BA) results are presented in the main text, while F1 scores are provided in Appendix \ref{sec:appendix:sim_add_results}. Additional results on the estimation of the number of node-level clusters are also provided in Appendix~\ref{sec:appendix:sim_add_results}.

\subsection{Simulation results: objective function behavior}\label{sec:simulation_obj}

\begin{figure}[!t]\centering
\includegraphics[width=1\columnwidth]{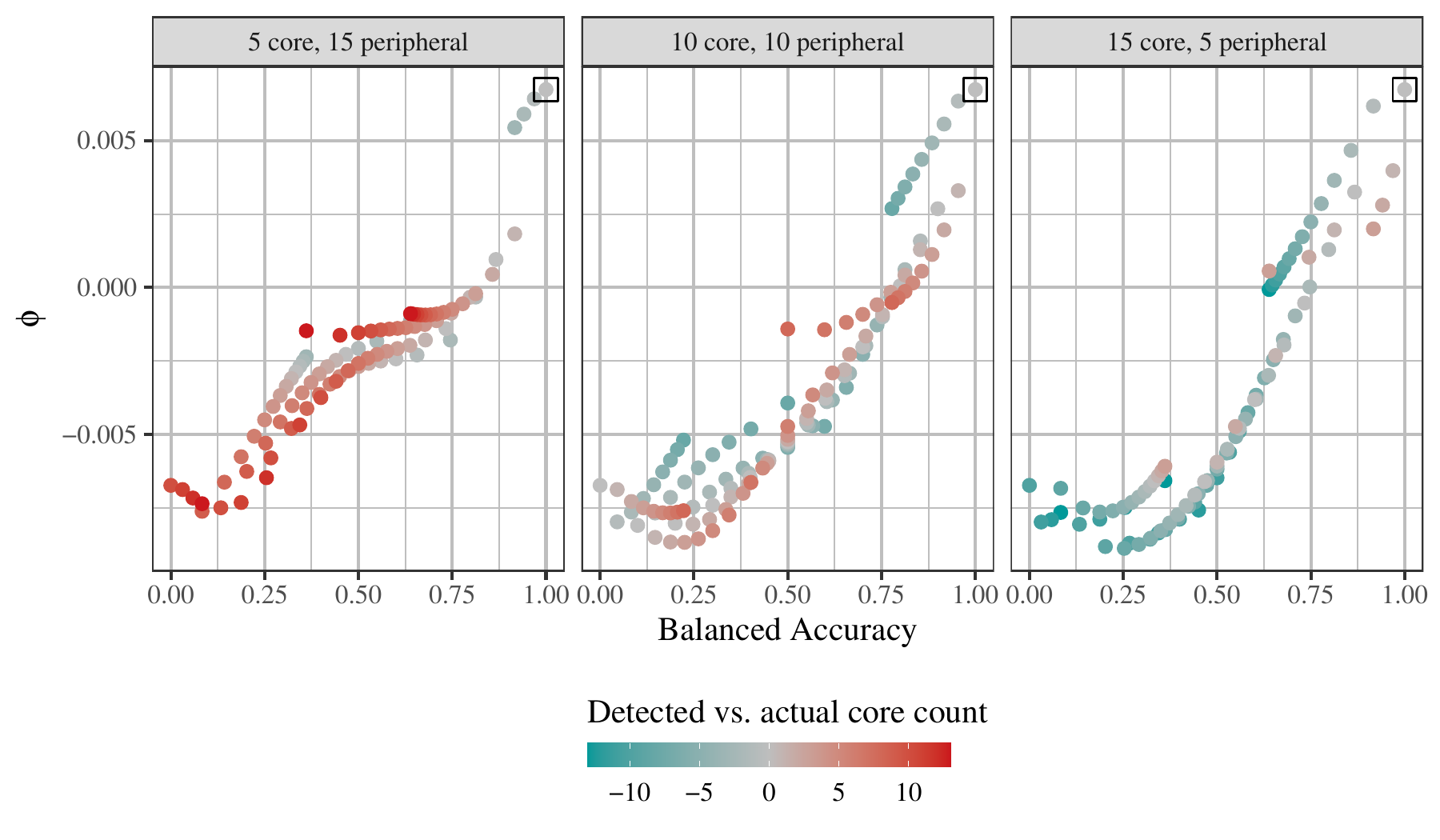}
\caption{\label{fig:fig2}  Scatter plot of objective function ($\phi$) vs. balanced accuracy (BA) scores for core-periphery networks with $n = 1000$ nodes, $K = 20$ communities, and core proportions of 25\%, 50\%, and 75\% (panel from left to right). Blue points represent solutions with more peripheral communities than the true structure, while red points indicate solutions with more core communities. The true solution ($\text{BA} = 1$) is highlighted with a square.}  
\end{figure}

 Figure~\ref{fig:fig2} shows the results of the evaluation of the behavior of the proposed objective function. The analysis is conducted on synthetic networks with $n = 1000$ nodes, $K = 20$ equally sized communities, and known core–periphery structures $\mathbf{z}$ under three different $\lambda$ values. Objective function values $\phi(\mathbf{z}; \bm{\Theta})$ are computed for all possible configurations of $\mathbf{z}$, using the actual data-generating $\bm{\Theta}$ values. There are 90 configurations when 5 communities are assigned to the core, 90 when 5 are assigned to the periphery, and 116 when 10 are assigned. 
 
 The BA measures the agreement between detected and actual $\mathbf{z}$. Blue points represent solutions with more peripheral communities than the true structure, while red points indicate solutions with more core communities. 
The true configuration ($\text{BA} = 1$) is highlighted with a square. The objective function $\phi$ reaches its maximum at the true solution, while close but suboptimal values are observed when one additional peripheral community is included in each of the three different configurations. 

\subsection{Simulation results: core-periphery detection performance}\label{sec:simulation_detection}
 
The second simulation study investigates how structural parameters, such as network size, number ($K$), and size of communities, and the proportion of core groups, influence the ability to detect core-periphery structures at the community level. 

\begin{figure}[!t]
    \centering
    \includegraphics[width=1\columnwidth]{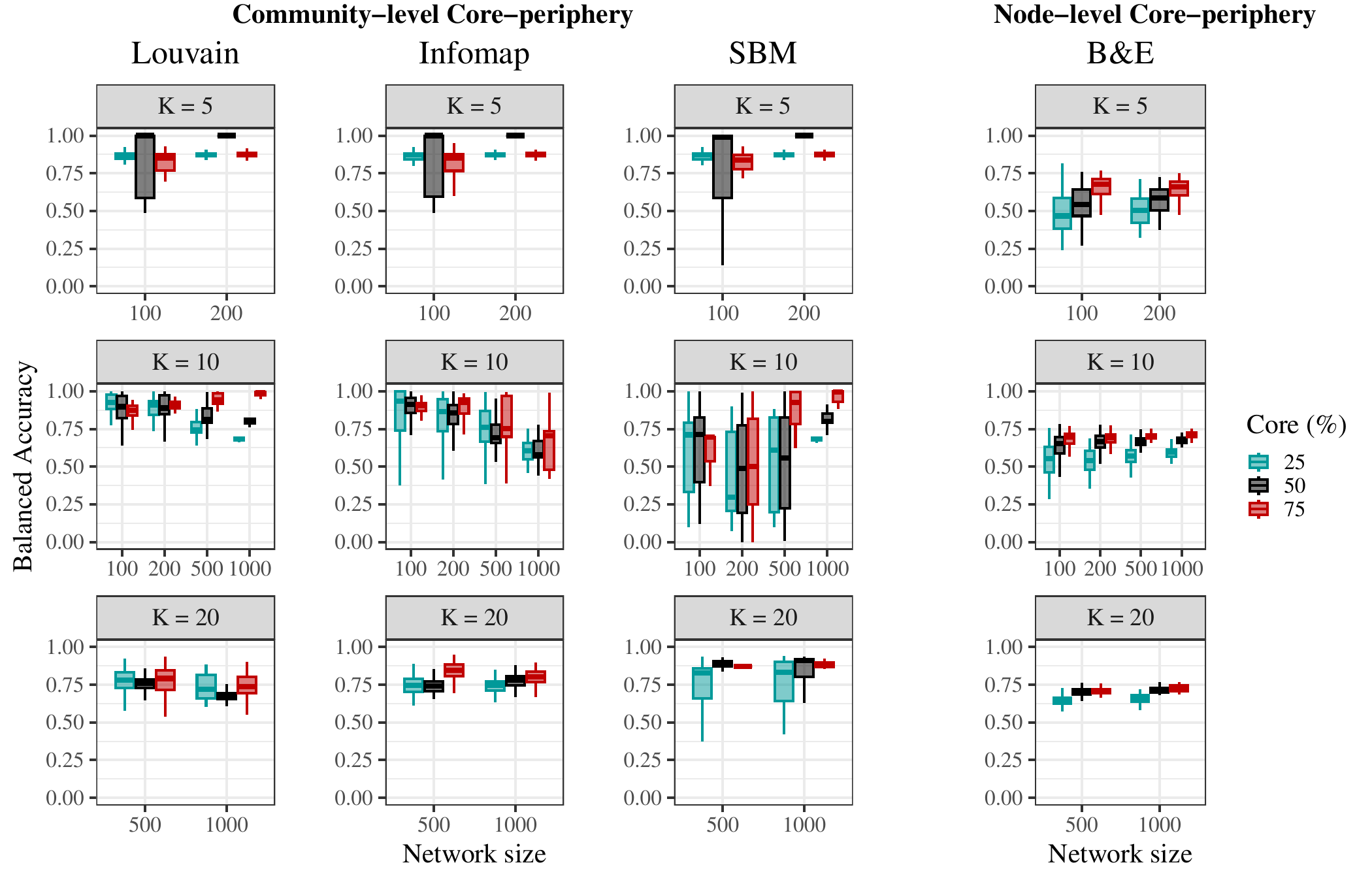}
    \caption{\textbf{Uniform} community sizes -- Balanced accuracy distribution across different network sizes and numbers of communities ($K$). 
    Results are shown for different community detection methods, including Louvain, Infomap, and SBM, as well as for the Borgatti and Everett (B\&E) approach.}
    \label{fig:fig3a}
\end{figure}

\begin{figure}[!t]
    \centering
    \includegraphics[width=1\columnwidth]{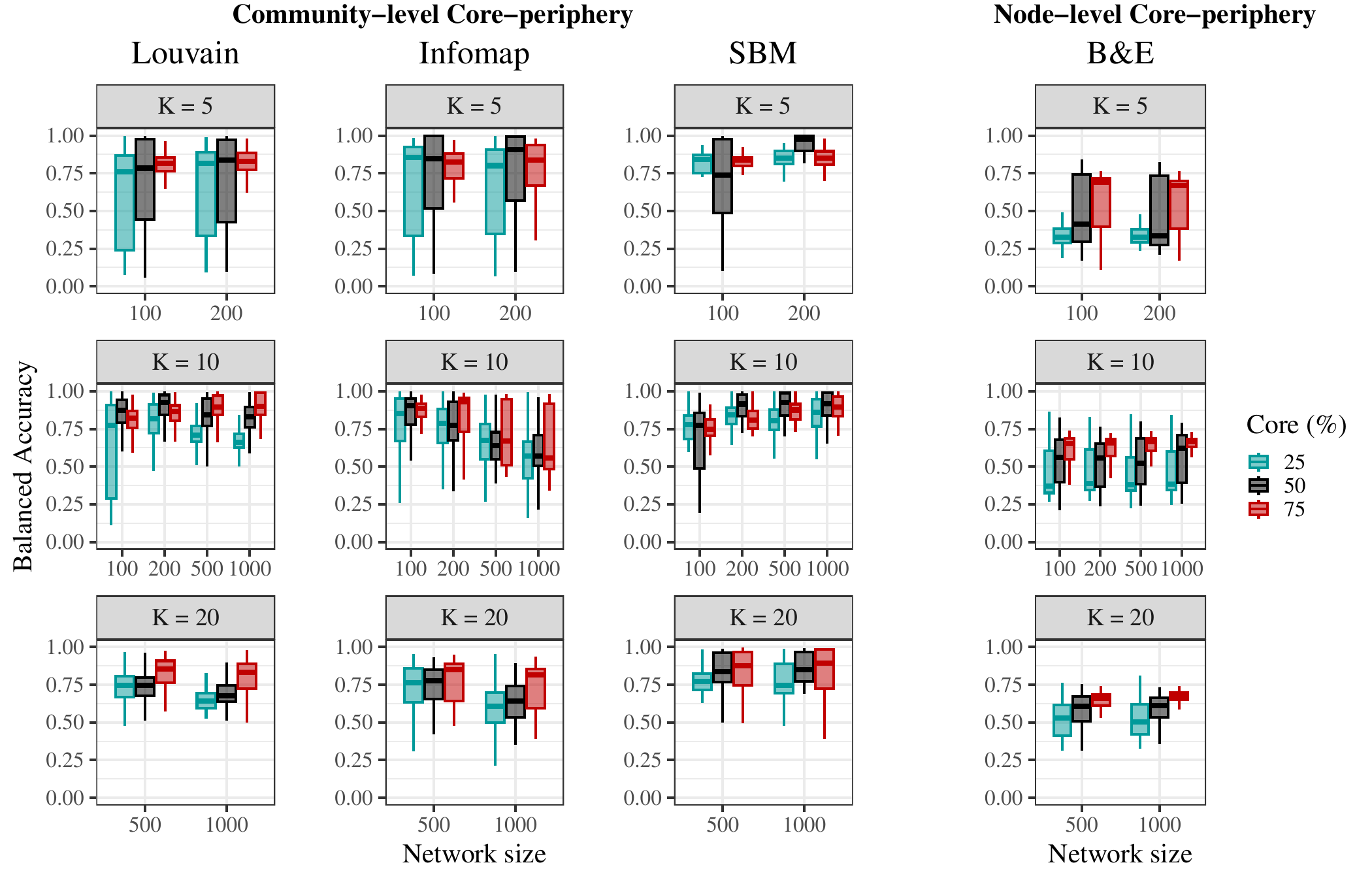}
    \caption{\textbf{Non-uniform} (Dirichlet-distributed) community sizes -- Balanced accuracy distribution across different network sizes and numbers of communities ($K$). 
    Results are shown for different community detection methods, including Louvain, Infomap, and SBM, as well as for the Borgatti and Everett (B\&E) approach.}
    \label{fig:fig3b}
\end{figure}

As shown in Figure~\ref{fig:fig3a} and \ref{fig:fig3b}, our method performs robustly across a wide range of realistic network configurations.
Since the approach assumes a fixed community partition as input, its performance depends on the quality of this initial grouping. To assess this, we apply the method following three standard node-level community detection algorithms: the binary SBM, implemented via the \textit{blockmodels} package \citep{Leger2021Blockmodels:Algorithm}, and the Louvain and Infomap algorithms, both available in the igraph package \citep{Csardi2025Network2.2.1,Csardi:igraph}. 

Overall, the proposed method demonstrates superior balanced accuracy for communities identified by the binary SBM compared with alternative approaches, particularly in networks with moderate to large sizes ($n = 500$ or $1000$). This is especially pronounced when the proportion of core communities is high (75\%) and the community sizes are non-uniform, reinforcing its suitability for complex and realistic network structures. When node-level community detection is implemented using Louvain, performance tends to be higher in smaller networks and under conditions of equal-sized communities.  When using Infomap for community detection, which emphasizes flow-based modular structures, the proposed approach tends to yield the lowest overall performance in large networks and in non-uniform scenarios. 

Across all algorithms, the proportion of core communities influences performance. In small networks, optimal performance is achieved at an intermediate core proportion (50\%), whereas in larger networks, performance peaks at 75\%. However, it is important to note that in scenarios with five communities, a core proportion of 25\% or 75\% results in only one core or one peripheral community, respectively. These are degenerate cases that pose a challenge for our method, which by design requires at least two communities in both the core and the periphery to define a valid core–periphery partition. Consequently, the method is not capable of identifying the true solutions in these settings, which explains the observed lower performance. Nevertheless, in small networks with equally sized communities, the method shows strong potential when used in combination with all three community detection algorithms.

For benchmarking purposes, we also compare our community-level approach with the node-level core–periphery detection model proposed by Borgatti and Everett, which remains a standard reference in this area. The method was implemented using the netUtils package in R \citep{Schoch2024CRAN:NetUtils}. As shown in Figure~\ref{fig:fig3a} and \ref{fig:fig3b}, the Borgatti–Everett model consistently achieves lower BA across all scenarios. This difference is particularly pronounced in networks with a low proportion of core communities, where the node-level approach fails to capture higher-order structural regularities. These results highlight the importance of leveraging community partitions to reliably identify community-level core–periphery organization at the meso-scale.

\section{Application: Italian academics co-authorship network}\label{sec:application}

The case study on co-authorship among Italian academic scholars illustrates the practical application of the proposed methodology. We examine two complementary settings. In the first, we infer node-level communities defined by high internal connectivity, with the objective of uncovering collaboration patterns at the researcher level. In the second, we assign community labels based on the regional affiliation of authors, allowing us to investigate collaboration dynamics and core–periphery structures across Italian regions. The underlying hypothesis is that geographical proximity may foster core regional clusters, and that the emergence of core and peripheral regions may further reflect thematic or institutional specialization within the academic system.

\subsection{Communities detected using SBM} 

We begin by applying the degree-corrected SBM \citep{Karrer2011StochasticNetworks} to the co-authorship network in order to obtain a meaningful community partition. This community detection approach is particularly suitable for this context, as it accounts for the heterogeneous degree distribution observed in the network, while still detecting assortative block structure. 

\begin{figure}[!t]
    \centering
    
    \begin{subfigure}[t]{.48\textwidth}
        \centering
        \includegraphics[width=\textwidth]{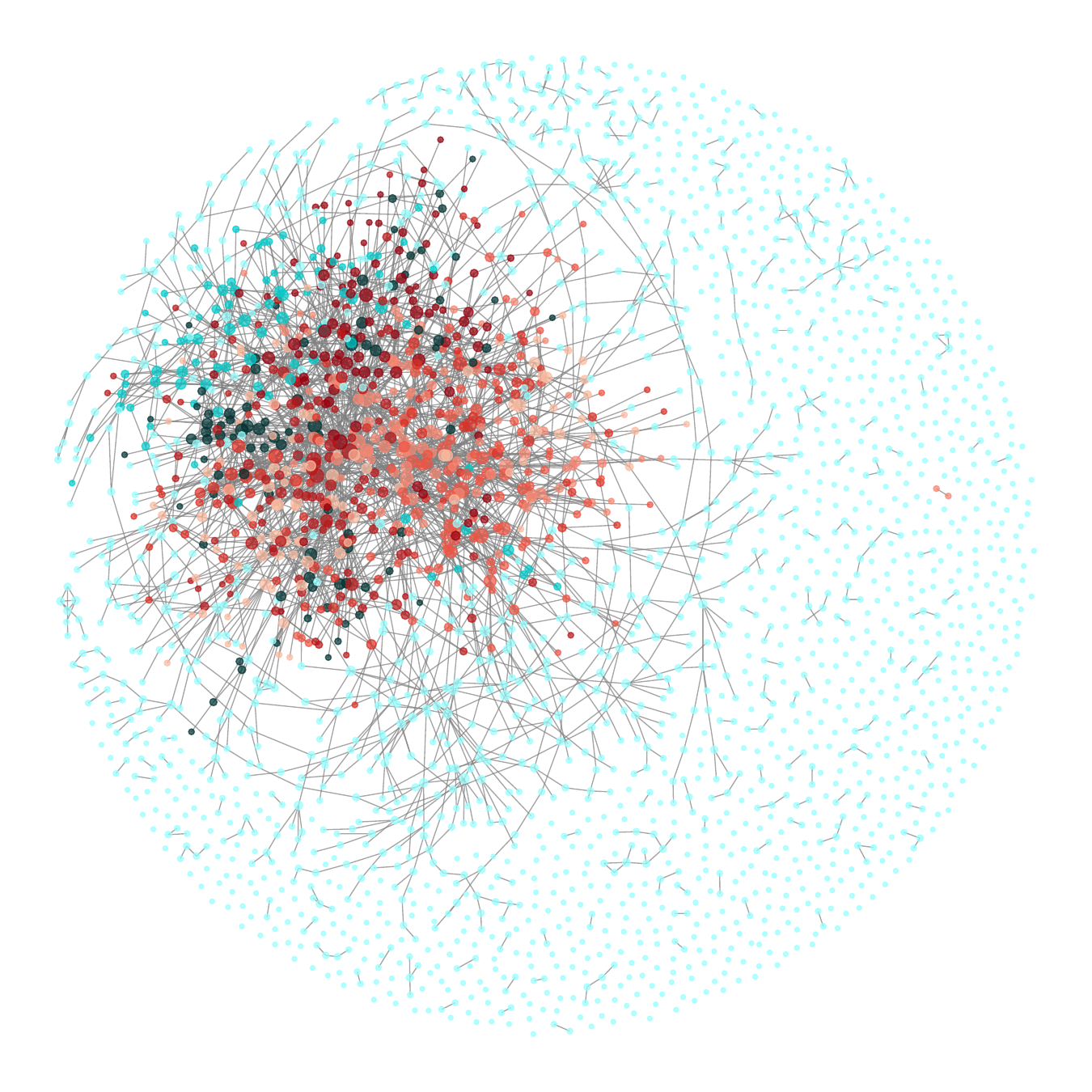}
        \caption{}
        \label{fig:sub6}
    \end{subfigure}
    \begin{subfigure}[t]{.51\textwidth}
        \centering
        \includegraphics[width=\textwidth]{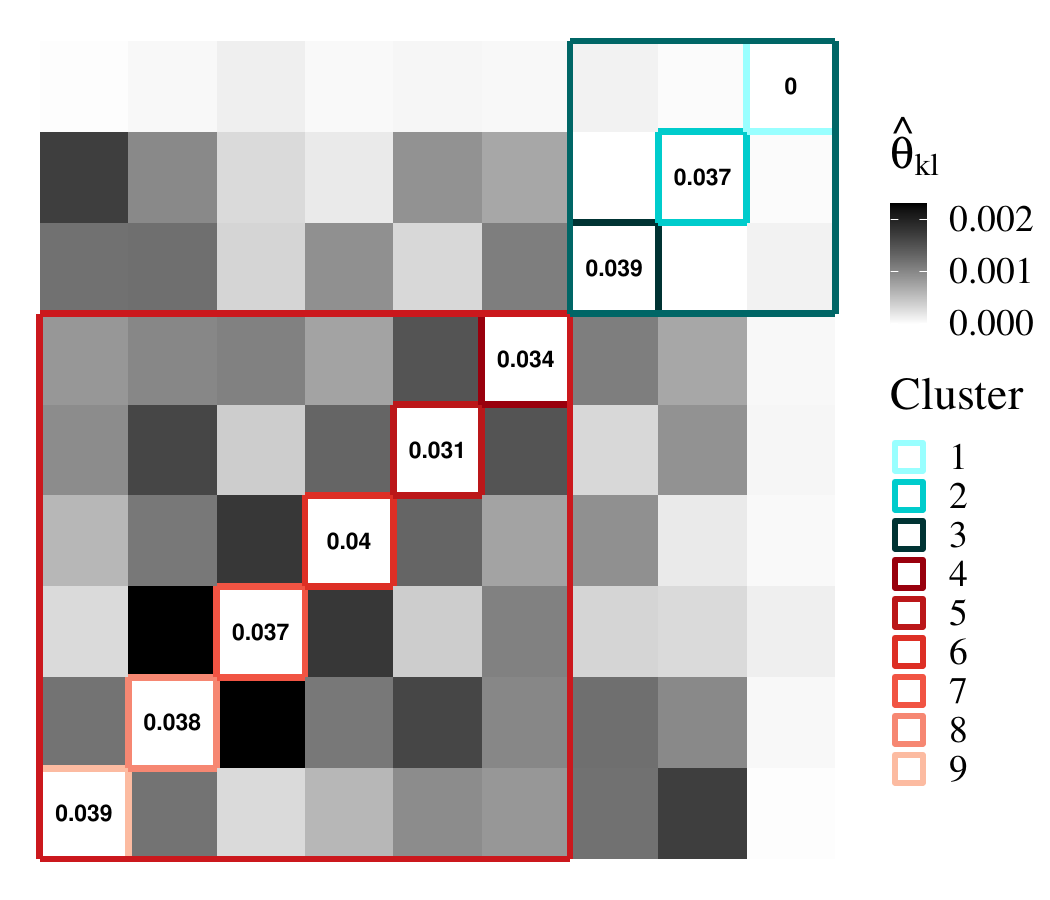}
        \caption{}
        \label{fig:sub7}
    \end{subfigure}

    \caption{(a) Co-authorship network partition in communities and core-periphery organization. (b) Co-authorship connectivity matrix ($\bm{\hat{\Theta}}$) representing the community-level core-periphery partition. Blue square denotes intra-periphery connectivity, red square intra-core connectivity.}
    \label{fig:fig4}
\end{figure}

The degree-corrected SBM identifies eight cohesive communities and one large group of sparsely connected or isolated nodes, which contains 70.6\% of all network nodes (light blue in Figure~\ref{fig:fig4}). Within this group, 66.9\% of nodes are disconnected from the largest component of the network, while only two interconnected but otherwise isolated nodes fall outside it. The distribution of the sizes of the remaining eight clusters is relatively uniform, ranging from 86 to 116 nodes.

From this community structure, we compute the estimated connectivity matrix $\hat{\bm{\Theta}}$ as described in Section~\ref{sec:method}, with entries capturing the edge density between communities. Using this matrix, our method evaluates all admissible partitions of the communities into core and periphery by optimizing the objective function $\phi(\mathbf{z}; \hat{\bm{\Theta}})$.
This procedure classifies the detected communities into six core and three peripheral groups. The resulting core set corresponds to densely connected and structurally homogeneous communities, whereas the peripheral set consists of groups characterized by sparse and heterogeneous connectivity patterns. Notably, the latter also includes the large light blue community, which contains the network's isolated nodes.

Results are presented in Figure~\ref{fig:fig4}. The left panel shows the community partition of the network, and the right panel displays the connectivity matrix $\hat{\bm{\Theta}}$ representing the community-level core-periphery partition. Each diagonal element of this matrix represents a community. 

To better reveal inter-regional links, we visualize the off-diagonal values in the $\hat{\bm{\Theta}}$ matrix, as a few highly assortative clusters produce dark diagonal cells that obscure off-diagonal values in the full heatmap.
Interaction probabilities between peripheral communities are nearly zero, as indicated by the white cells in the blue square (intra-periphery block), and their links to core communities are weak. In contrast, core communities exhibit more dense, though still relatively weak, interactions, represented by the light grey cells in the red square (intra-core block). 

For comparison, we also applied the node-level core–periphery detection model of \cite{Borgatti2000ModelsStructures}. This approach results in a trivial partition of the network into high- and low-degree nodes, essentially reproducing degree heterogeneity: low-degree nodes naturally appear as peripheral (Table \ref{tab:tab2}). To assess the consistency between our community-level core–periphery assignment and the node-level B\&E partition, we compared the two partitions on the co-authorship network. The results show a BA of 0.78 and an F1 of 0.89, indicating partial agreement between the two approaches but also meaningful differences in how the ``core'' role is defined. Most nodes identified as peripheral by our method (2,016 out of 2,483) are also classified as peripheral by the B\&E model, while 67 are labelled as core. Conversely, among the nodes belonging to core communities in our framework, 190 (31\%) are also assigned to the B\&E core, whereas 418 (69\%) are classified as peripheral. This pattern suggests that our method identifies additional locally central actors—those having dense connections within a core community—who may not appear as globally central in the B\&E sense, consistent with the community-level focus of our approach.

\begin{table}[!t]
\centering
\caption{Degree distribution by core membership assigned using Borgatti and Everett model.}
\label{tab:tab2}
\begin{tabular}{ccccccc}
\toprule
 & $n$ & Mean & SD & Median & Min & Max \\
\midrule
Periphery & 2434 & 1.53 & 1.66 & 1 & 0 & 6 \\
Core & 257  & 9.96 & 3.66 & 9 & 7 & 30 \\
\bottomrule
\end{tabular}
\end{table}

The distinction between core and periphery structure may be explained by thematic specialization. To characterize the thematic content of each community and of the core and periphery, we combine term frequency (tf) and term frequency-inverse document frequency (tf–idf) statistics \cite{Robinson2017TextR}, computed from words appearing in titles and keywords, with data on the frequency of publication venue (scientific journals). These quantities were computed in R using the functions implemented in the tidytext package \citep{Robinson2025Text0.4.3}. The tf results are presented in Figure~\ref{fig:fig5} in the main text, while tf–idf results can be found in Figure~\ref{appx:fig4} of Appendix~\ref{sec:appendix:tfidf}.

Standard text preprocessing steps were applied to ensure that only representative words are retained. Specifically, we split the texts into individual words (word tokenization) and lemmatized them to reduce them to their unique dictionary form. Next, we removed stop-words (i.e., prevalent and non-informative words such as ``and'', ``the'', and ``to''), numbers, and symbols. The data were further cleaned to exclude extremely common but non-informative words, such as ``data'', ``model'', ``analysis'', and ``approach''.

The largest peripheral cluster (Cluster 1) is dominated by themes from the social sciences and applied fields, as reflected in frequent terms such as ``social'', ``innovation'', and ``management'', and in high tf–idf terms such as ``democracy'', ``populism'', and ``luxury''. It is also associated with various Italian sociological journals, including {\em Italian Sociological Review}, {\em Partecipazione e Conflitto}, and {\em Italian Journal of Sociology of Education}. Furthermore, a substantial proportion of the publications in this cluster appeared in more recent years, suggesting a relatively young and still consolidating research area, characterized by limited collaborations.

The second peripheral cluster (Cluster 2) centers on a highly specialized domain of statistical theory and methodology. Distinctive terms (``bayesian'', ``nonparametrics'', ``dirichlet'', ``pitman'', ``gibbs'') and high-impact journals (i.e. {\em Biometrika}, {\em Electronic Journal of Statistics}, and {\em Journal of the American Statistical Association}) indicate that this cluster represents advanced knowledge production in complex statistical methodology.
Its theoretical orientation and narrow focus further characterize it as peripheral in the overall collaboration structure.

The third peripheral group (Cluster 3) integrates applied statistics with a wide range of application areas. Dominant tf terms include ``ordinal'', ``effect'', ``covariance'', and ``country'', while tf–idf highlights terms such as ``symbolic'', ``interlock'', ``histogram'', and ``port''. Journals like {\em Electronic Journal of Applied Statistical Analysis}, {\em Quality and Quantity}, as well as field-specific outlets such as {\em Social Indicators  Research}, {\em British Food Journal}, and {\em Public Health Nutrition}, position this cluster within applied quantitative methods for social and economic research, with a notable subfield focusing on nutrition and food related health studies. The peripheral role of the cluster reflects its predominantly applied nature and cross-disciplinary involvement. Rather than forming integrated collaborative groups, they contribute analytical expertise in different thematic areas, giving rise to weaker and more transient collaborations.

Overall, the three peripheral clusters comprise scholars from distinct scientific domains, each focused on research topics that frequently require collaboration with researchers possessing complementary—rather than thematically aligned—expertise. Based on the journal distribution reported in Fig. \ref{fig:fig6}, Cluster 1 is predominantly composed of sociologists and economists working on themes such as sustainability, political studies, urban and regional sociology, and education, and exhibits a generally low propensity to collaborate. Clusters 2 and 3 comprise two distinct profiles of statisticians whose reciprocal collaboration networks are comparatively limited: the former due to their focus on applied research within specialized domains, and the latter because of their strong orientation toward theoretical topics. As a result, they tend to collaborate mainly with statisticians located in the core communities. For example, members of Cluster 2 often work with environmental and medical researchers (mainly located in the core Cluster 9 as described below) rather than extensively within their own clusters.

These peripheral communities are the only groups in the network that display almost no interaction with one another.

In contrast, the six core clusters are characterized by greater thematic homogeneity.
Cluster 4 centers on methodological innovation in time-series analysis, and financial and economic modeling, featuring terms such as ``time series'', ``autocovariance'', ``stationarity''). Cluster 5 emphasizes general statistical modeling and inference, with distinctive terms including ``em algorithm'', ``mixture'', ``symmetric'', ``invariance'', ``irt''.
Clusters 6 and 7 are strongly disciplinary. Cluster 6, linked to journals such as {\em Economic Modelling} and {\em Regional Science and Urban Economics}, is oriented toward land use and environmental studies (frequent terms include ``land'', ``urban'', ``spatial'', ``sprawl'', ``degradation'', ``desertification''). In contrast, Cluster 7 is rooted in demography and family studies, featuring distinctive terms such as ``birth rate'', ``fertility'', ``childlessness'', and ``grandchild'', and characterized by journals including {\em Demographic Research} and {\em European Journal of Population}. Cluster 8 partially overlaps thematically with Cluster 7, with terms like ``small area'', ``index'', ``poverty'', and ``parturition'', and is characterized by journals such as the {\em Journal of Official Statistics} and {\em Health Policy}. Cluster 9 combines applied statistics with biomedical and environmental applications (reflected in terms such as ``disease'', ``effect'', ``asthma'', ``seismicity'') with a strong link to environmental and health research, represented by journals such as {\em Ecological Indicators}, {\em Statistics in Medicine}, and {\em Science of the Total Environment}. 
The members of the clusters detected as core tend to collaborate repeatedly within well-established research domains, forming dense, stable ties that sustain the network’s structural cohesion.

\begin{figure}[!t]
    \centering
    
    \begin{subfigure}[t]{.495\textwidth}
        \centering
        \includegraphics[width=\textwidth]{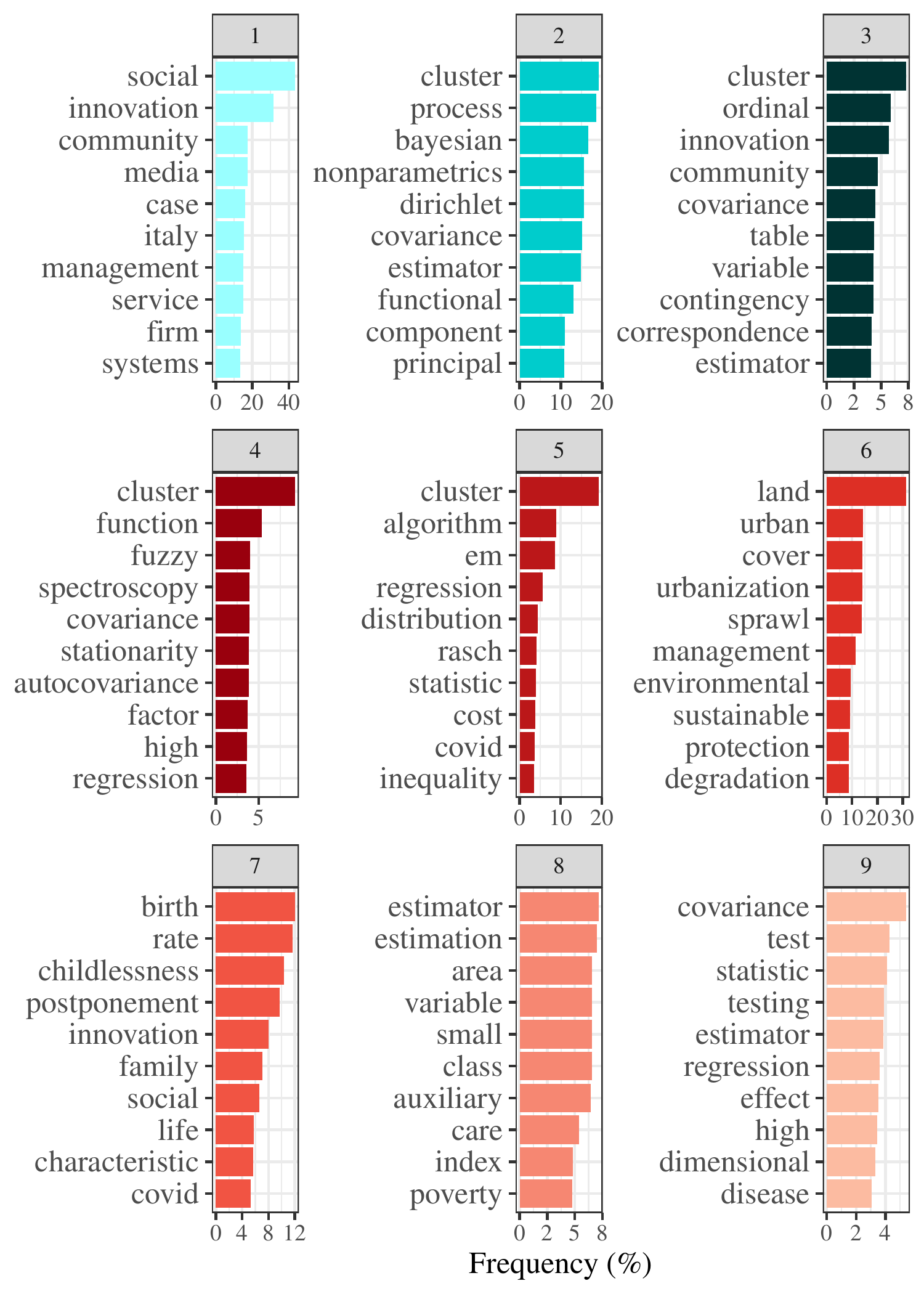}
        \caption{Paper topic}
        \label{fig:sub8}
    \end{subfigure}
    \begin{subfigure}[t]{.495\textwidth}
        \centering
        \includegraphics[width=\textwidth]{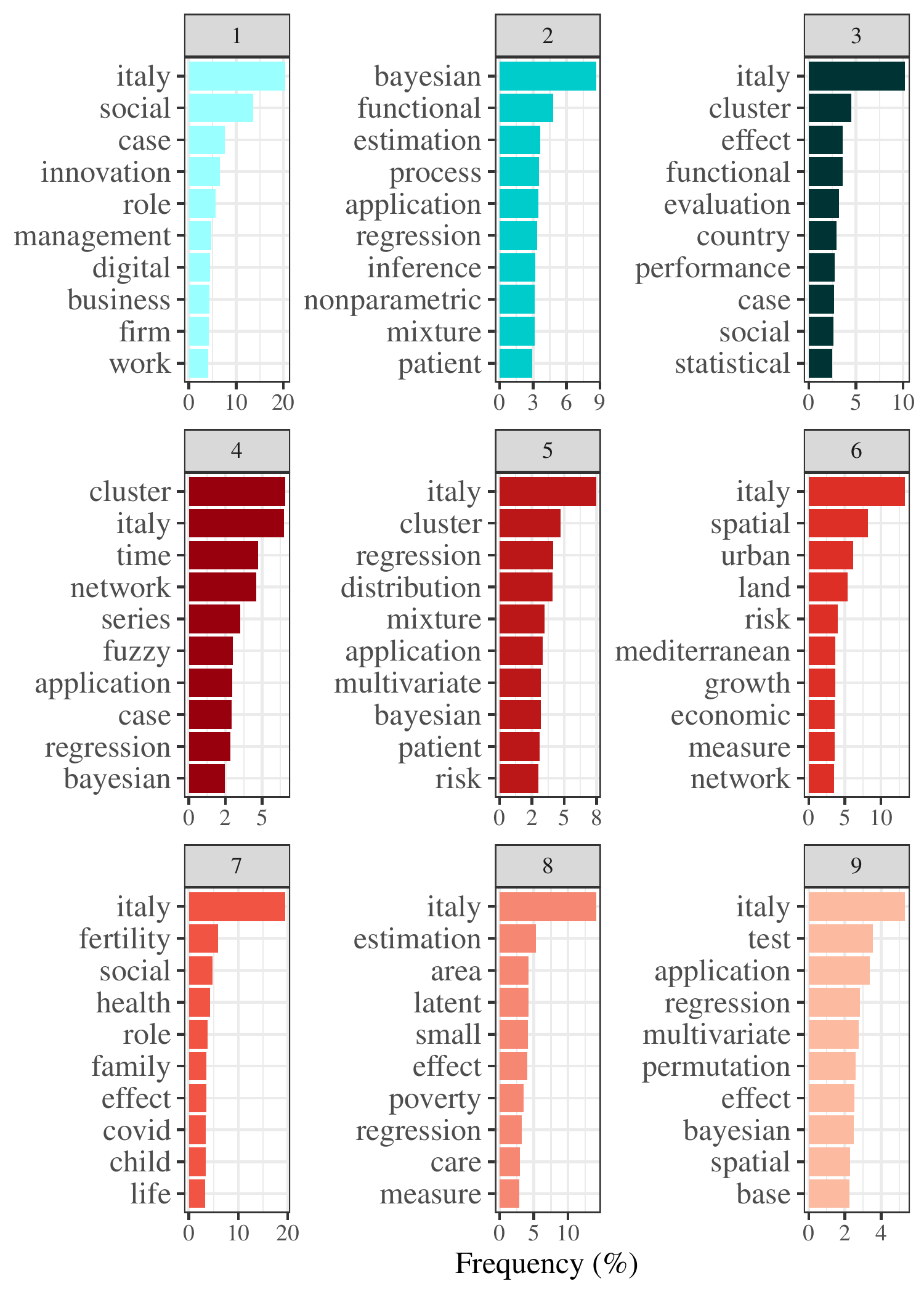}
        \caption{Paper title}
        \label{fig:sub9}
    \end{subfigure}

    \caption{(a) Distribution of topic term frequency (tf) by cluster and core-periphery organization. (b) Distribution of title term frequency (tf) by cluster and core-periphery organization.}
    \label{fig:fig5}
\end{figure}

\begin{figure}[t!]
    \centering
    
        \includegraphics[width=\textwidth]{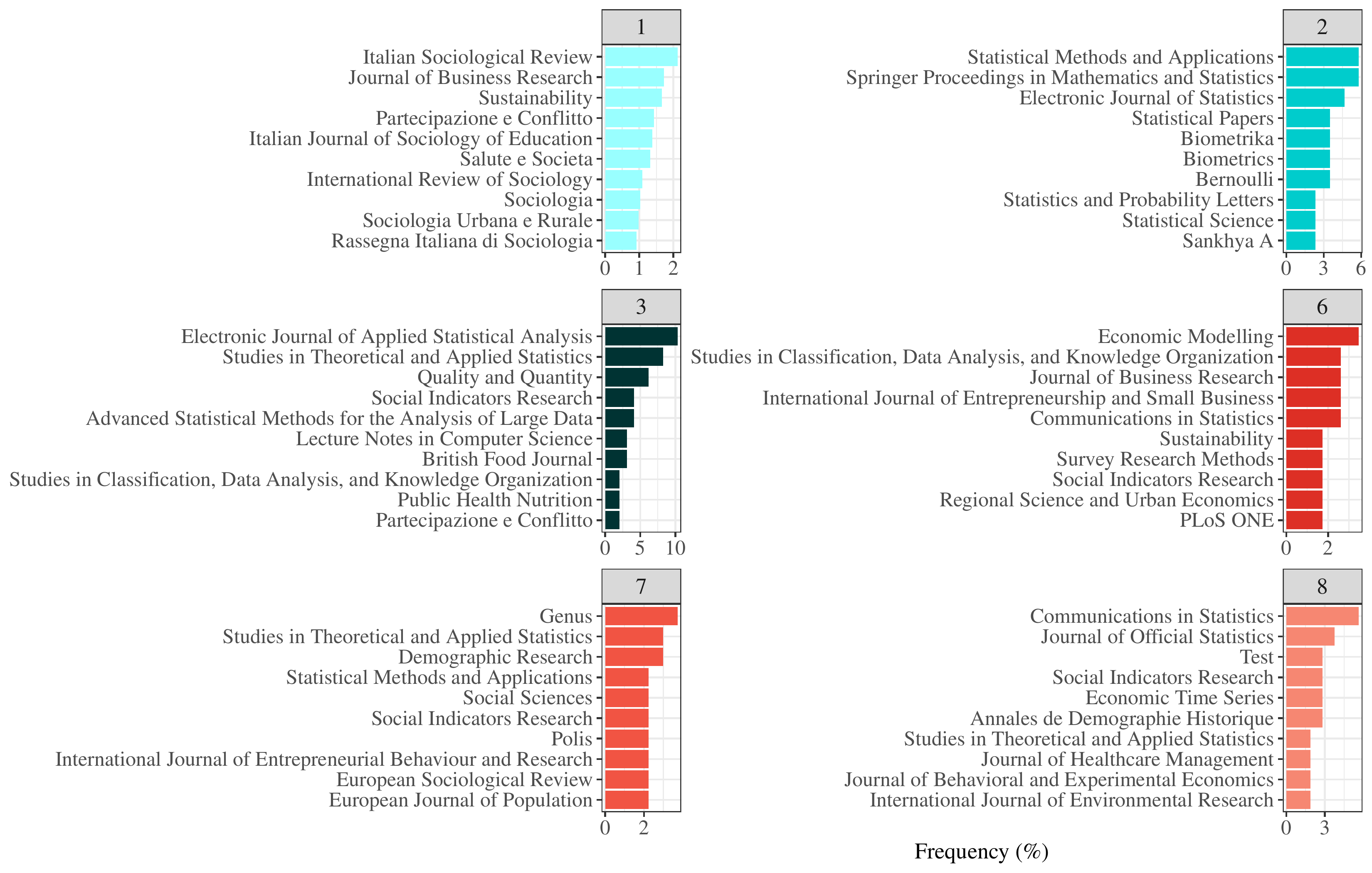}

    \caption{Distribution of journal frequency by cluster}
    \label{fig:fig6}
\end{figure}

\begin{figure}[t!]
    \centering
    
    \begin{subfigure}[t]{.49\textwidth}
        \centering
        \includegraphics[width=\textwidth]{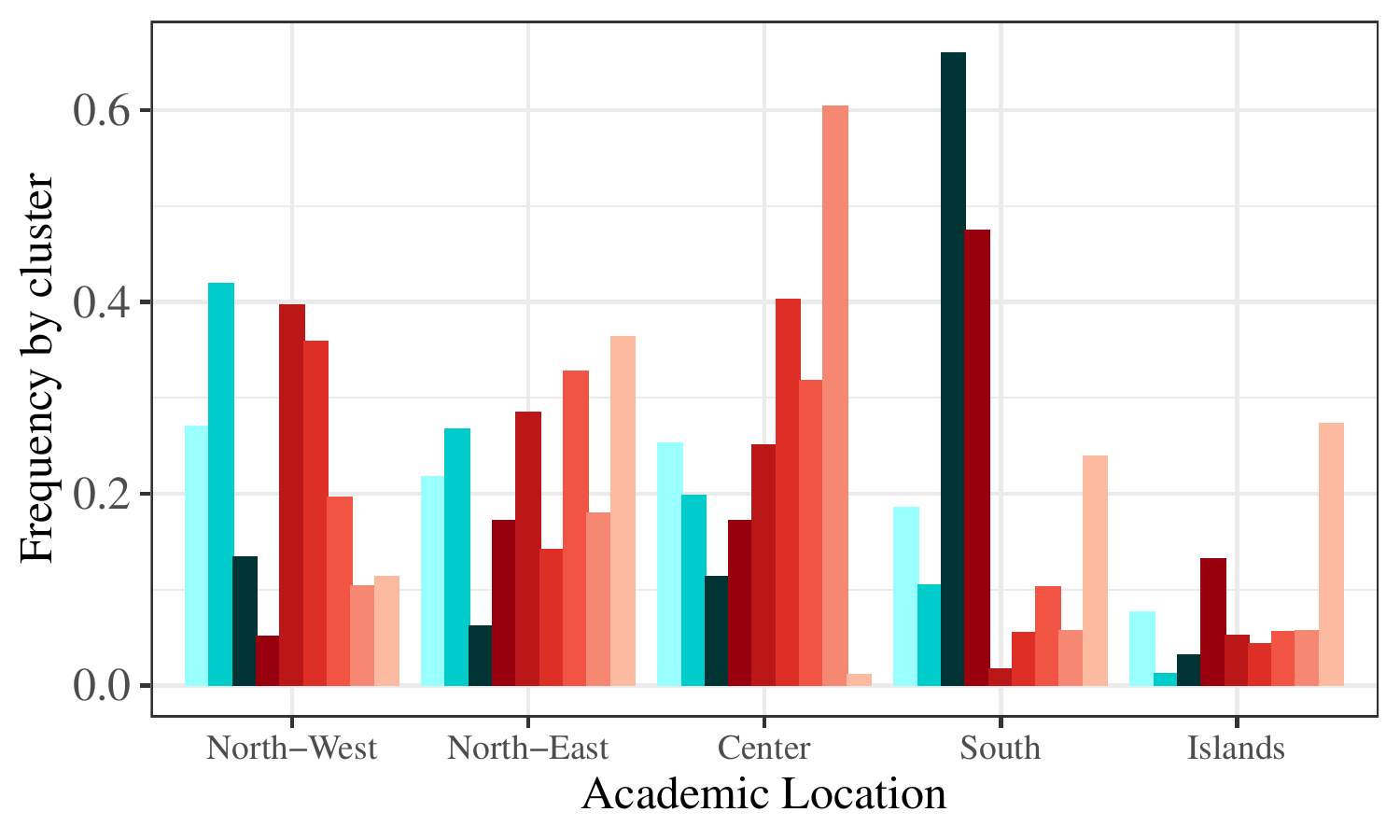}
        \caption{}
        \label{fig:sub10}
    \end{subfigure}
    \begin{subfigure}[t]{.49\textwidth}
        \centering
        \includegraphics[width=\textwidth]{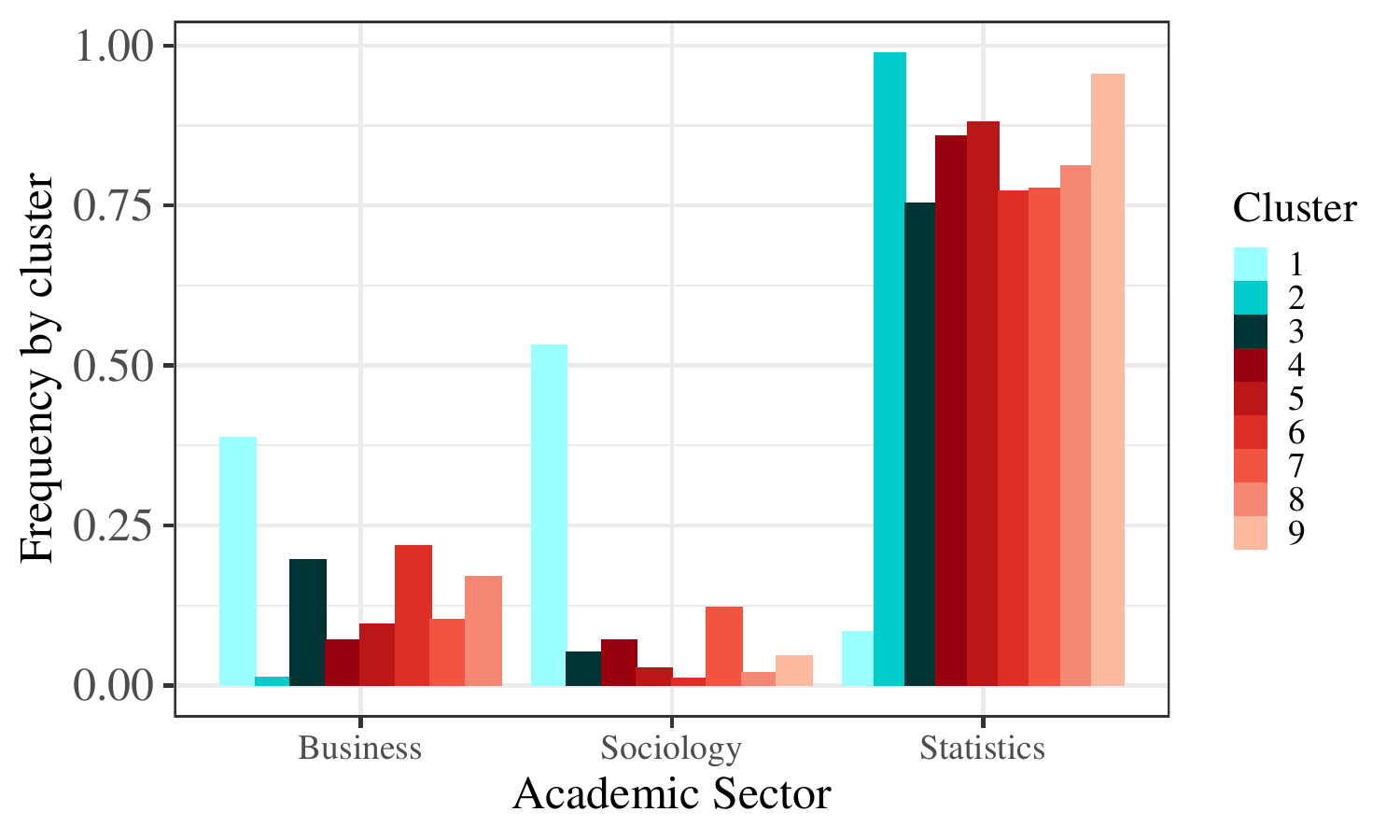}
        \caption{}
        \label{fig:sub11}
    \end{subfigure}

    \caption{(a) Distribution of authors’ geographic locations by cluster. (b) Distribution of authors’ academic sectors by cluster.}
    \label{fig:fig7}
\end{figure}

Figure~\ref{fig:fig7} presents information on researcher characteristics, including the distribution of authors’ geographic locations and authors’ academic sectors by cluster. Peripheral cluster 3 shows a marked predominance of researchers based in southern Italy (66.0\%), compared with a much smaller southern presence in the core clusters, where it rarely exceeds 20\% (Figure~\ref{fig:sub10}). Cluster 2 is dominated by academics from the North-East and North-West, clearly associated with the Bayesian non-parametric groups centered at the University of Padova and Bocconi University. 

As expected from exploratory analyses, the large peripheral Cluster 1 groups together most of the non-statisticians, who are only weakly connected to the rest of the network (Figure~\ref{fig:sub11}). 91.7\% of the nodes in that community belong to the business or social sciences, highlighting the bridging role of statistics as a multidisciplinary field. 
Notably, the overall core-periphery connectivity is guided by statisticians. This high propensity to collaborate is in line with previous studies on the Italian Statistician community \citep{bacci2023insights,deStefano2023quality}.

Overall, these results confirm that peripheral clusters not only diverge thematically but also have distinctive social and institutional compositions.

\begin{figure}
    \centering
    \includegraphics[width=0.5\linewidth]{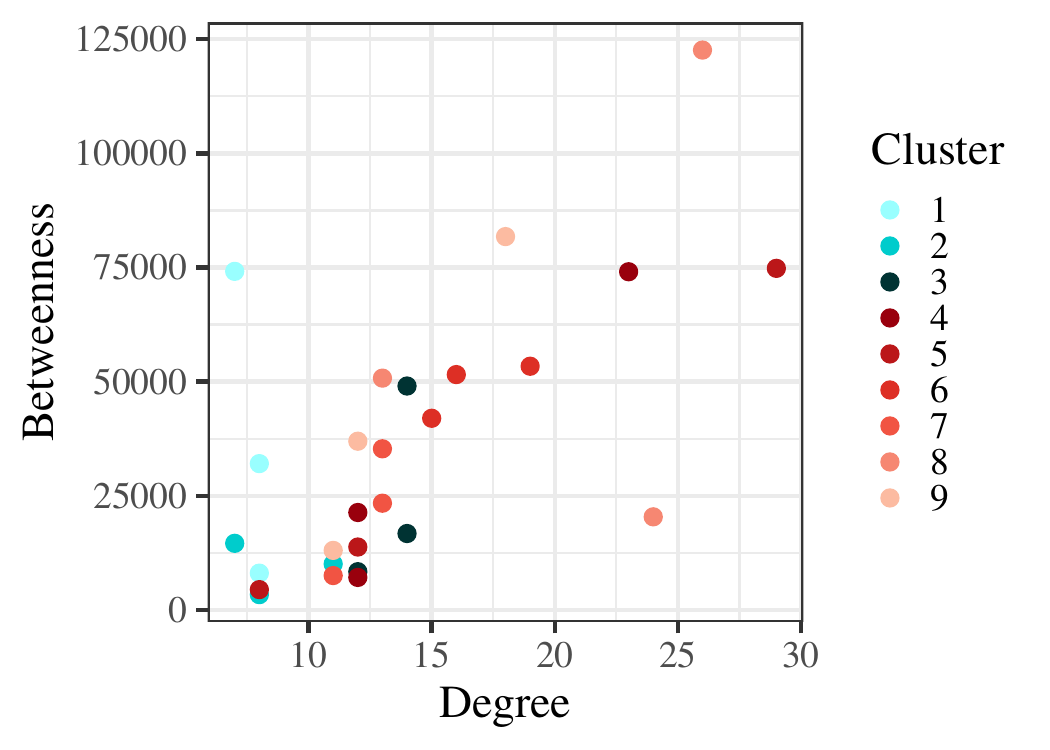}
    \caption{Top three leaders' degree and betweenness centrality for core and peripheral clusters.}
    \label{fig:fig8}
\end{figure}

To further characterize and interpret the core and periphery groups, we identified leaders within each community as the three members with the highest number of inter-community collaborations (a total of 27 leaders, 9 periphery leaders, and 18 core leaders). In the case of ties, degree centrality and then betweenness centrality were used as tie-breakers. Core leaders consistently display higher values across all centrality measures, reflecting their more central and well-connected positions in the collaboration network (Figure~\ref{fig:fig8}). On average, their degree is 15.9, compared to 9.9 for peripheral leaders, and their mean betweenness centrality reaches 40,800, nearly double that of peripheral leaders (24,100), indicating a greater role in bridging collaborations across communities.

Leaders are predominantly full professors, concentrated in central and northern regions such as Lazio, Lombardia, and Toscana, and are exclusively affiliated with the Statistics sector. Periphery leaders generally exhibit a more heterogeneous mix of academic ranks (including researchers and associates), and a broader geographic distribution, extending to southern regions such as Campania, Puglia, and Calabria. The only community led by scholars outside the field of statistics (i.e., sociology) is peripheral. 

\begin{table}[b!]
    \centering
\begin{tabular}{llrr}
  \toprule
   &  & \makecell{Periphery} & \makecell{Core} \\    
      \cmidrule{1-4} 
      
     Total& & 9 & 18 \\    
      \cmidrule{1-4}
      
Field 
 & Statistics & 7 & 18 \\ 
 & Business & 0 & 0 \\ 
 & Sociology & 2 & 0 \\ 
   \cmidrule{1-4}     
Role  
 & Full Professor & 4 &11 \\ 
 & Associate & 4 & 5 \\ 
 & Researcher & 1 & 2 \\ 
     \cmidrule{1-4} 
Location 
   & North-West & 0 & 5 \\ 
   & North-East & 3 & 1 \\ 
   & Center & 1 & 8 \\    
   & South & 5 & 3 \\ 
   & Islands & 0 & 1 \\ 
   \bottomrule
\end{tabular}
    \caption{Frequency of leaders by sector, role, and location across core and peripheral clusters.}
    \label{tab:tab3}
\end{table}

\subsection{Communities defined by geographical regions}

Using the categorization of the 20 Italian regions as the node-level community partition, the estimated average connection probability is 0.027 within clusters and 0.00039 between clusters, revealing a strongly assortative regional structure. The proposed method identifies 14 core and 6 peripheral regions: Basilicata, Friuli Venezia Giulia, Liguria, Molise, Puglia, and Valle d’Aosta. Highly represented regions (over 250 authors) are all assigned to the core, whereas those scarcely represented (fewer than 15 authors) are in the periphery. One determinant of the core-periphery organization is the geographical proximity, and our method allows us to detect which regions have few or no connections reciprocally.

\begin{figure}[t!]
    \centering
    
       \includegraphics[width=\textwidth]{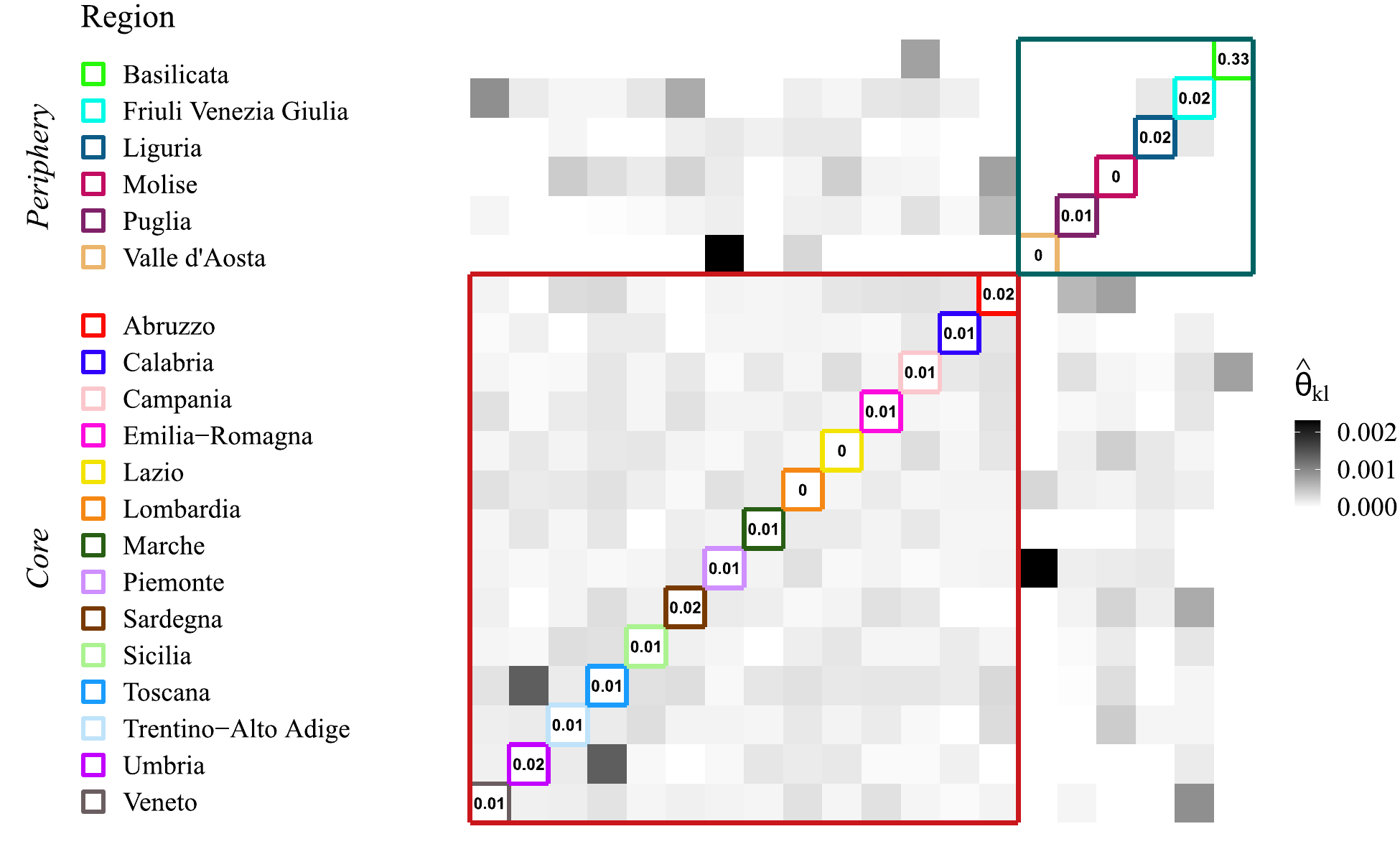}

    \caption{Regional co-authorship connectivity matrix ($\hat{\Theta}$) illustrating the core-periphery partition of the Italian regions. Blue square denotes intra-periphery connectivity, red square intra-core connectivity.}
    \label{fig:fig9}
\end{figure}

Figure~\ref{fig:fig9} shows very weak intra-periphery interactions (top-right blue square) and sparser core–periphery links compared to core–core links (bottom-left red square). 

For each region, the three leaders were identified as outlined above (a total of 59 leaders, 17 representing peripheral regions and 42 core regions). 
Core regional leaders show higher centrality values, with an average degree of 11.62 compared to 5 for peripheral region leaders and an average betweenness of 26584.79 compared to 4252.81 for peripheral region leaders, indicating their role as connectors in the collaboration network. They are predominantly full and associate professors in the Statistics sector. In contrast, peripheral regional leaders have a greater representation of researchers and associates, and show greater disciplinary diversity, with some leaders affiliated with the business field (e.g., all leaders from Basilicata and Valle d'Aosta). 

\begin{table}[b!]
    \centering
\begin{tabular}{llrr}
  \toprule
   &  & \makecell{Periphery} & \makecell{Core} \\    
      \cmidrule{1-4} 
      
     Total& & 17 & 42 \\    
      \cmidrule{1-4}
      
Field 
 & Statistics & 12 & 40 \\ 
 & Business & 5 & 0 \\ 
 & Sociology & 0 & 2 \\ 
   \cmidrule{1-4}     
Role  
 & Full Professor & 6 & 23 \\ 
 & Associate & 7 & 15 \\ 
 & Researcher & 4 & 4 \\ 
   \bottomrule
\end{tabular}
    \caption{Frequency of regional leaders by sector and role across core and peripheral clusters.}
    \label{tab:tab4}
\end{table}

\section{Discussion and Future Research}\label{sec:discussion}

We introduced a community-level approach to core–periphery detection, designed to distinguish densely connected core groups from sparsely connected peripheral ones based on both the density and strength of inter-community interactions. By formulating an objective function tailored to this task, our method provides a scalable, interpretable, and flexible framework applicable to a variety of real-world networks. The proposed methodology is showcased to analyze a collaboration network of Italian academic scholars.

Our results highlight how the proposed framework departs from classical node-level models (e.g., \cite{Borgatti2000ModelsStructures}) by reconceptualizing the periphery. Rather than representing a residual set of loosely connected nodes, the periphery may also include cohesive and internally dense communities that remain weakly integrated into the broader network, tending to form their ties primarily with the core. This insight helps explain why peripheries can be both intellectually rich and structurally isolated, as shown in our empirical application to the Italian academic collaboration network. 

The findings underscore the potential of community-level core–periphery analysis to support science policy. Identifying peripheral groups and their leaders can inform initiatives to strengthen inter-community collaborations, broaden participation across geographic and disciplinary boundaries, and promote a more balanced distribution of resources or enhance scientific innovation, fostering, for instance, preriphery-to-periphery collaboration. More broadly, mapping nested structures enriches our understanding of how influence, knowledge flows, and collaboration dynamics are organized at the group level, complementing established node-level approaches \citep{Zelnio2012IdentifyingScience, Karlovcec2016Core-peripherySlovenia, Sedita2020TheLiterature, Wedell2022CenterperipheryCommunities}.

From a methodological point of view, an interesting avenue for future extension concerns the choice of summary statistics in the objective function. Our formulation uses mean-based measures, which guarantee a single optimum and stable results. Using more robust alternatives, such as the median, could produce multiple optima, reducing stability but potentially increasing robustness to asymmetric connection densities, where mean-based measures are sensitive to outliers. For a brief discussion and an example, see Appendix~\ref{sec:appendix:median}.

At the same time, several limitations point toward directions for future work.  
First, in the current framework, clustering and core–periphery detection are performed sequentially. While this modularity simplifies computation and interpretation, it makes the core-periphery detection dependent on the initial node-level partition. A natural extension is the development of a nested SBM capable of jointly estimating node-level communities and community-level core–periphery roles, thereby integrating the two levels into a unified inferential framework \citep{oto2019BayesianBlockmodeling, Come2021HierarchicalLikelihood}.  
Second, categorical attributes (e.g., role, research interest, location) are used mainly to interpret core–periphery roles and, when treated as community labels, to assess how different group types align with core or peripheral positions.  Future work should embed both categorical and continuous attributes directly in the detection step, for example through attributed network clustering algorithms \citep{Zhang2023CommunityStatisticians} or attributed block models \citep{Stanley2019StochasticAttributes}, to enable a richer account of how social and institutional factors shape structural positions.  
Third, our empirical analysis focuses on a specific small academic community in Italy, constrained by data availability. The method should be tested on larger, more diverse scientific fields and extended to temporal networks \citep{Balland2019Network20032017}, to capture how core–periphery roles evolve as collaborations expand and reorganize.
Lastly, the application to the co-authorship network considered only binary connections, disregarding edge weights and thereby omitting potentially useful information about collaboration intensity. The proposed framework can, however, be readily extended to weighted networks, particularly in the node-level community detection phase. For example, existing approaches such as weighted SBMs or modularity-based methods for weighted networks can be directly incorporated into our framework \citep{Fortunato2016CommunityGuide,Ng2021WeightedModel}.
Extending the core–periphery detection step to weighted settings would require adapting the objective function to account for weighted connectivity patterns \citep{Borgatti2000ModelsStructures,Tudisco2019AGraphs,Rombach2012Core-PeripheryNetworks}. Nevertheless, this remains a more challenging task, as weight distributions in empirical networks are often highly skewed, over- or under-dispersed, and influenced by exogenous factors, such as visibility, team size, or resource availability, rather than purely structural position.

In summary, the present work provides the first explicit framework for detecting nested core–periphery structures at the community level. Future extensions will broaden its scope and deepen its explanatory power, contributing to the broader study of hierarchical organization in complex networks.

\subsection*{Acknowledgements}

Sara Geremia and Domenico De Stefano acknowledge financial support under the National Recovery and Resilience Plan (NRRP), with European Union resources, NextGeneration EU - National Recovery and Resilience Plan, Mission 4 - Component 1 - Investment 4.1 - Project Title Models and methods for the analysis of collaboration networks – CUP J53D23011540006.\\

The doctoral scholarship is co-financed with European Union resources, NextGeneration EU - National Recovery and Resilience Plan, Mission 4 - Component 1 - Investment 4.1 – CUP J92B22000900007. \\

The authors thank Agenzia Lavoro \& SviluppoImpresa – Friuli Venezia Giulia for the support.

\clearpage

\bibliographystyle{apalike}
\bibliography{references}

\clearpage

\appendix

\section{Details of data generation in simulation study}\label{sec:appendix:sim_details}

\setcounter{table}{0}
\renewcommand{\thetable}{A\arabic{table}}

Different network sizes, $n = \{100, 200, 500, 1000 \}$ and numbers of communities, 
$K = \{5, 10, 15 \}$ are considered. Each node is assigned to one of these communities 
according to a probability distribution $\boldsymbol{\tau} = (\tau_1, \ldots, \tau_k, \ldots, \tau_K )$:

\begin{itemize}
    \item \textbf{Uniform}

    Each community is equally likely, so
    $$\tau_k = \frac{1}{K}, \qquad k = 1,\ldots,K,$$

    \item \textbf{Non-uniform}

    The community proportions are drawn from a symmetric Dirichlet distribution:
    $$\boldsymbol{\tau} \sim \mathrm{Dirichlet}(\mathbf{1}_K).$$
\end{itemize}

Node labels are then drawn as
    $$q_i \mid \boldsymbol{\tau} \sim \mathrm{Categorical}(\boldsymbol{\tau}).$$

A parameter $\lambda = \{25, 50, 75\}$ controls the percentage of communities belonging to the core. 

Networks are generated according to a SBM whose connectivity probability parameters are defined as a function of $n$ (number of nodes) and community and core-periphery membership. Some considerations concerning the definition of these connectivity probabilities. In sparse SBMs, considering an expected node degree of at least $\log(n)$ can be relevant for ensuring connectivity and community structure detection, particularly when connection probabilities scale as $\log(n) / n$.
In line with this theory, all our connection probabilities are a function of $\log(n) / n$, so that the graph remains sparse as $n \rightarrow \infty$. These probabilities are adjusted by a scaling factor (tuned by means of empirical evaluation and sensitivity analysis) which controls the relative density of connections between different types of node pairs (Table \ref{appx:tab1}), allowing the generation of networks with both community and core-periphery structures.

\begin{table}[h!]
\centering
\begin{tabular}{l c}
\hline
\textbf{Edge type} & \textbf{Connection probability} \\
\hline
Within-community & $8 \cdot \log(n)/n$ \\

Core--core & $\log(n)/n$ \\

Core--periphery & $(1/8)\cdot \log(n)/n$ \\

Periphery--periphery & $(1/40) \cdot \log(n)/n$ \\
\hline
\end{tabular}
\caption{Scaling of edge probabilities by edge type.}
\label{appx:tab1}
\end{table}

\section{Alternative summary statistics}\label{sec:appendix:median}

\setcounter{figure}{0}
\renewcommand{\thefigure}{B\arabic{figure}}

An alternative specification of the objective function in Equation~\ref{eq:objective} replaces the mean edge frequencies with the median edge frequencies of the core and periphery, denoted with $m_c$ and $m_p$ and replaces the standard deviations with the interquartile ranges $IQR_c$ and $IQR_p$. The resulting alternative objective function is
\begin{equation}
    \tilde{\phi}(\mathbf{z};\hat{\bm{\Theta}}) = \big(\delta_c + m_c - IQR_c\big) - \big(\delta_p + m_p + IQR_p\big).
\end{equation}

\begin{figure}[h]\centering
\includegraphics[width=1\columnwidth]{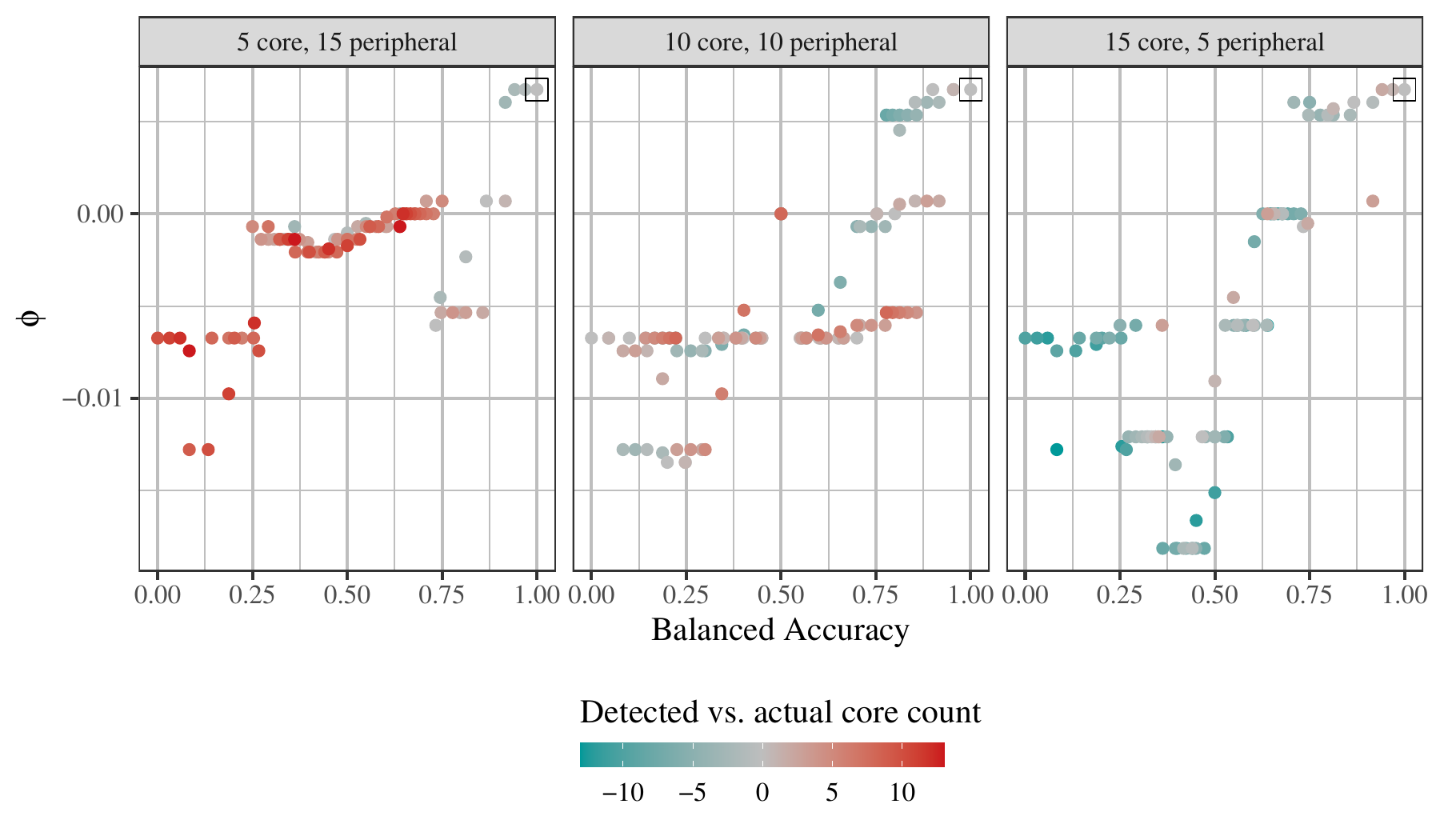}
\caption{\label{appx:fig1}  Scatter plot of the alternative objective function $\tilde{\phi}(\mathbf{z};\hat{\bm{\Theta}})$ using median and IQR vs. balanced accuracy for core-periphery networks with $n = 1000$ nodes, $K = 20$ communities, and core proportions of 25\%, 50\%, and 75\%. Blue points represent solutions with more peripheral communities than the true structure, while red points indicate solutions with more core communities. The true solution ($\text{Balanced Accuracy} = 1$) is highlighted with a square.}  
\end{figure}

Figure~\ref{appx:fig1} reports the values of $\tilde{\phi}(\mathbf{z};\hat{\bm{\Theta}})$ for the simulation study of Section~\ref{sec:simulation_obj}. While using more robust statistics such as medians and interquartile ranges may improve robustness to skewed or asymmetric connection densities, their discrete nature can create flat regions in the objective surface and generate multiple local optima. As a result, optimization becomes less stable, potentially complicating the identification of a unique community-level core–periphery partition.

\section{Additional results of simulation study}\label{sec:appendix:sim_add_results}

\setcounter{figure}{0}
\renewcommand{\thefigure}{C\arabic{figure}}

\begin{figure}[t!]
    \centering
    
    \begin{subfigure}[t]{1\textwidth}
        \centering
        \includegraphics[width=.75\columnwidth]{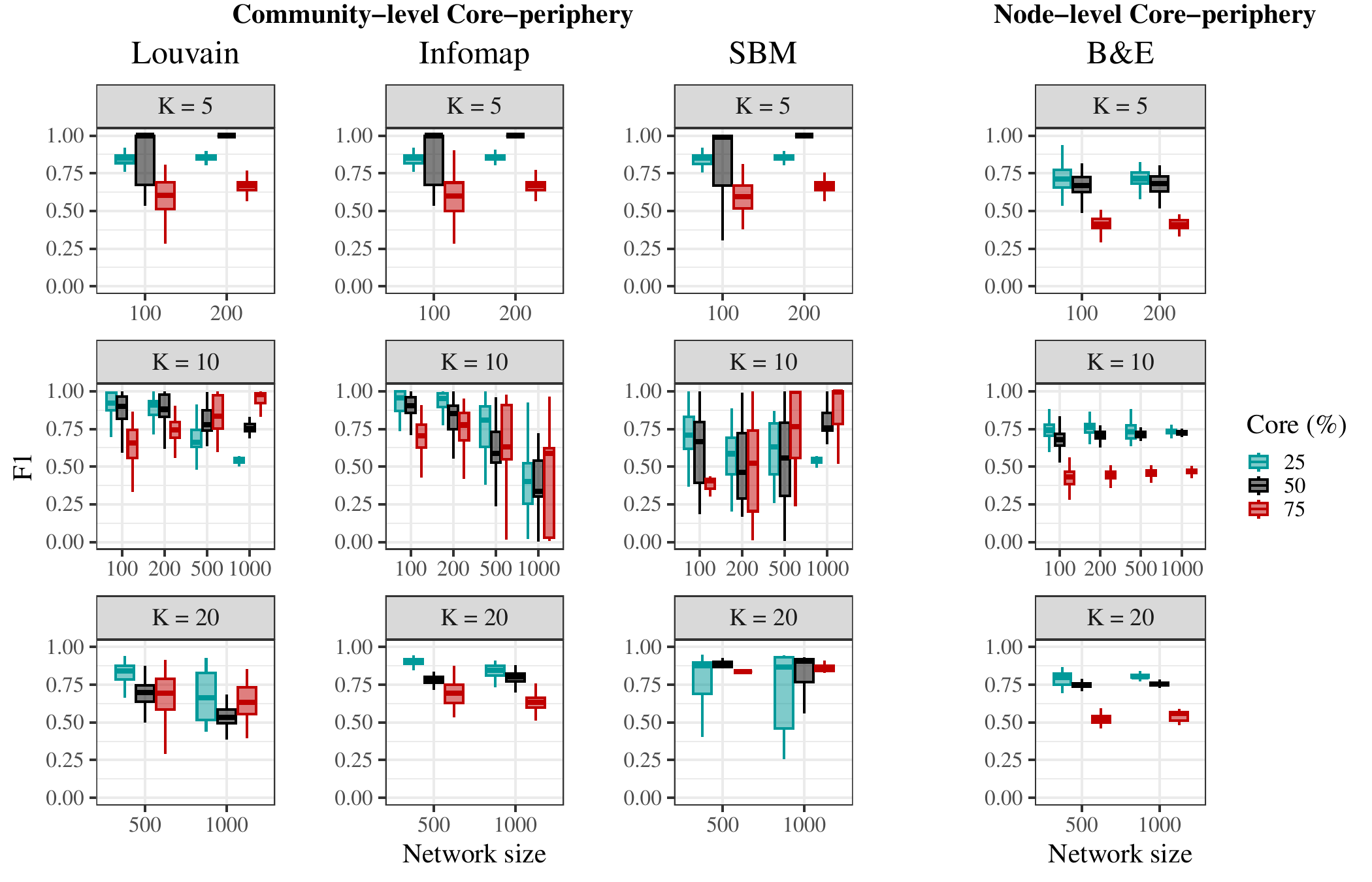}
        \caption{Uniform}
        \label{appx:sub3}
    \end{subfigure}
    \hfill
    \begin{subfigure}[t]{1\textwidth}
        \centering
        \includegraphics[width=.75\columnwidth]{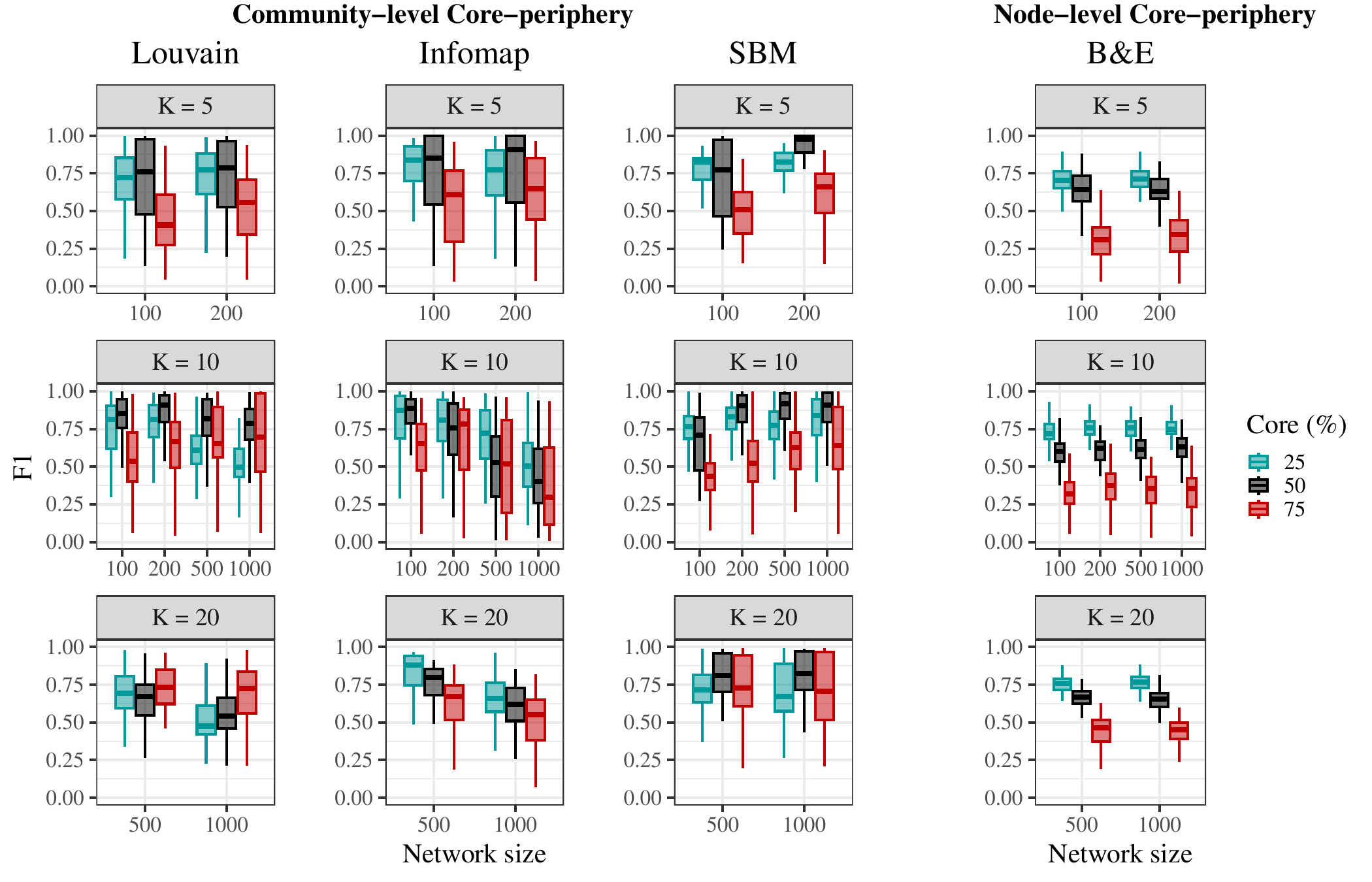}
        \caption{Non-uniform}
        \label{appx:sub4}
    \end{subfigure}

\caption{(a) F1 score distribution across different network sizes ($n$) and numbers of communities ($K$) for uniform community sizes. (b) F1 score distribution across different network sizes ($n$) and numbers of communities ($K$) for non-uniform (Dirichlet-distributed) community sizes.}
    \label{appx:fig3}
\end{figure}

Figure \ref{appx:fig3} reports F1 score values for the simulation study of Section~\ref{sec:simulation_detection}. Results with regards to F1 scores are in line with the BA results commented in the main text.

\begin{figure}[t!]
    \centering
    
    \begin{subfigure}[t]{1\textwidth}
        \centering
        \includegraphics[width=.75\columnwidth]{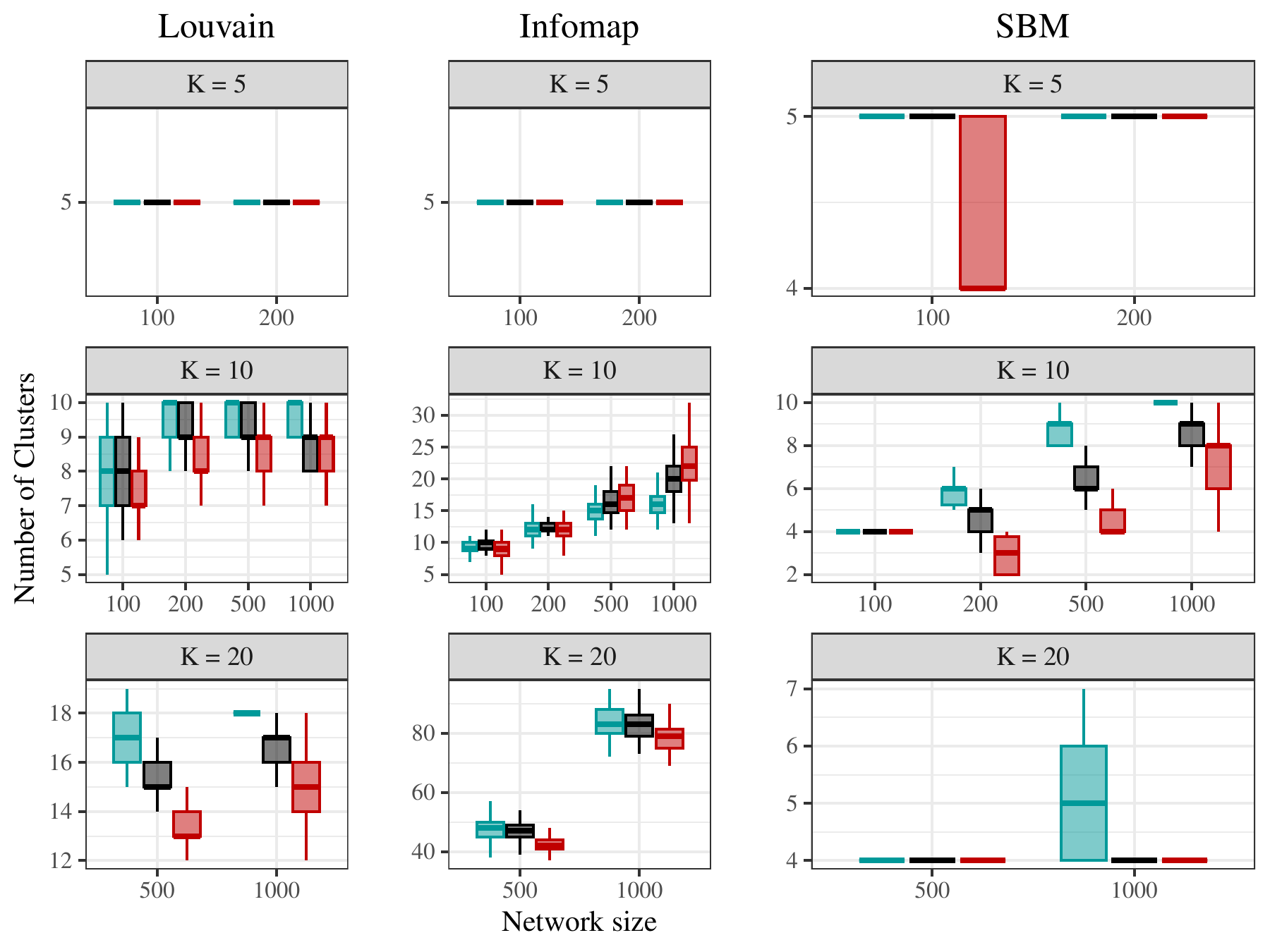}
        \caption{Uniform}
        \label{appx:sub1}
    \end{subfigure}
    \hfill
    \begin{subfigure}[t]{1\textwidth}
        \centering
        \includegraphics[width=.75\columnwidth]{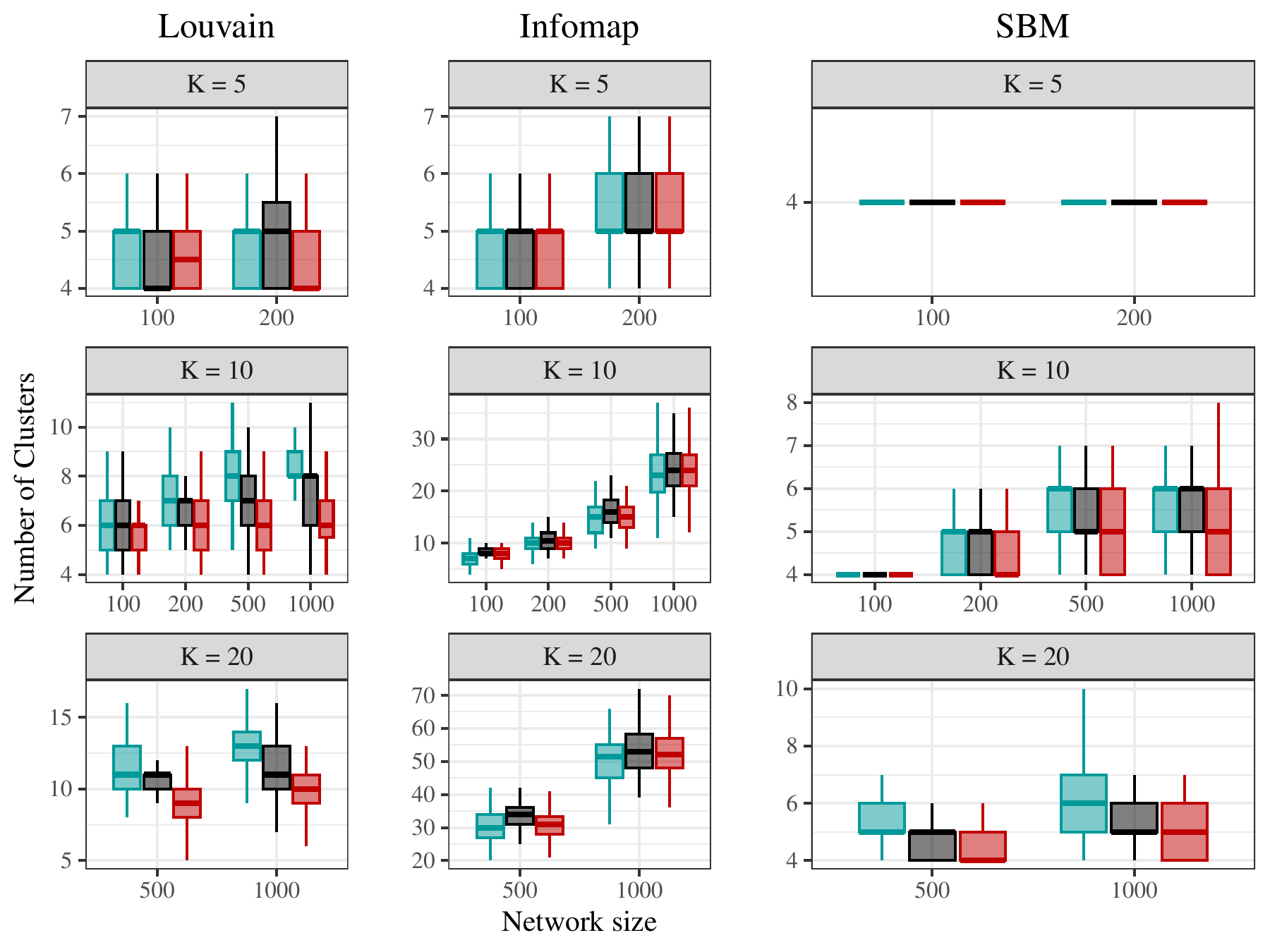}
        \caption{Non-uniform}
        \label{appx:sub2}
    \end{subfigure}

\caption{(a) Estimated number of clusters distribution across different network sizes ($n$) and numbers of communities ($K$) for uniform community sizes. (b) Estimated number of clusters distribution across different network sizes ($n$) and numbers of communities ($K$) for non-uniform (Dirichlet-distributed) community sizes.}
    \label{appx:fig2}
\end{figure}

Figure~\ref{appx:fig2} shows the estimated number of clusters for each node-level community detection method (Louvain, Infomap, and SBM) in the simulation study of Section~\ref{sec:simulation_detection}. In scenarios where the data-generating value of $K$ is large, all three methods struggle to recover the true number of communities: Louvain and SBM typically underestimate the number of clusters, whereas Infomap tends to overestimate it. Despite these inaccuracies, the results for core–periphery identification remain satisfactory (see Figures~\ref{fig:fig3a} and~\ref{fig:fig3b} in the main text). In particular, although the node-level SBM frequently underestimates $K$, the method nevertheless identifies the core–periphery partition correctly and consistently. This suggests that, when the number of clusters is underestimated, the estimated node-level communities tend to merge true communities within the core, preserving the higher-order core–periphery structure even if the finer-grained clustering is not fully recovered.

\section{tf-idf distributions}\label{sec:appendix:tfidf}

\setcounter{figure}{0}
\renewcommand{\thefigure}{D\arabic{figure}}

\begin{figure}[t!]
    \centering
    
    \begin{subfigure}[t]{.495\textwidth}
        \centering
        \includegraphics[width=\textwidth]{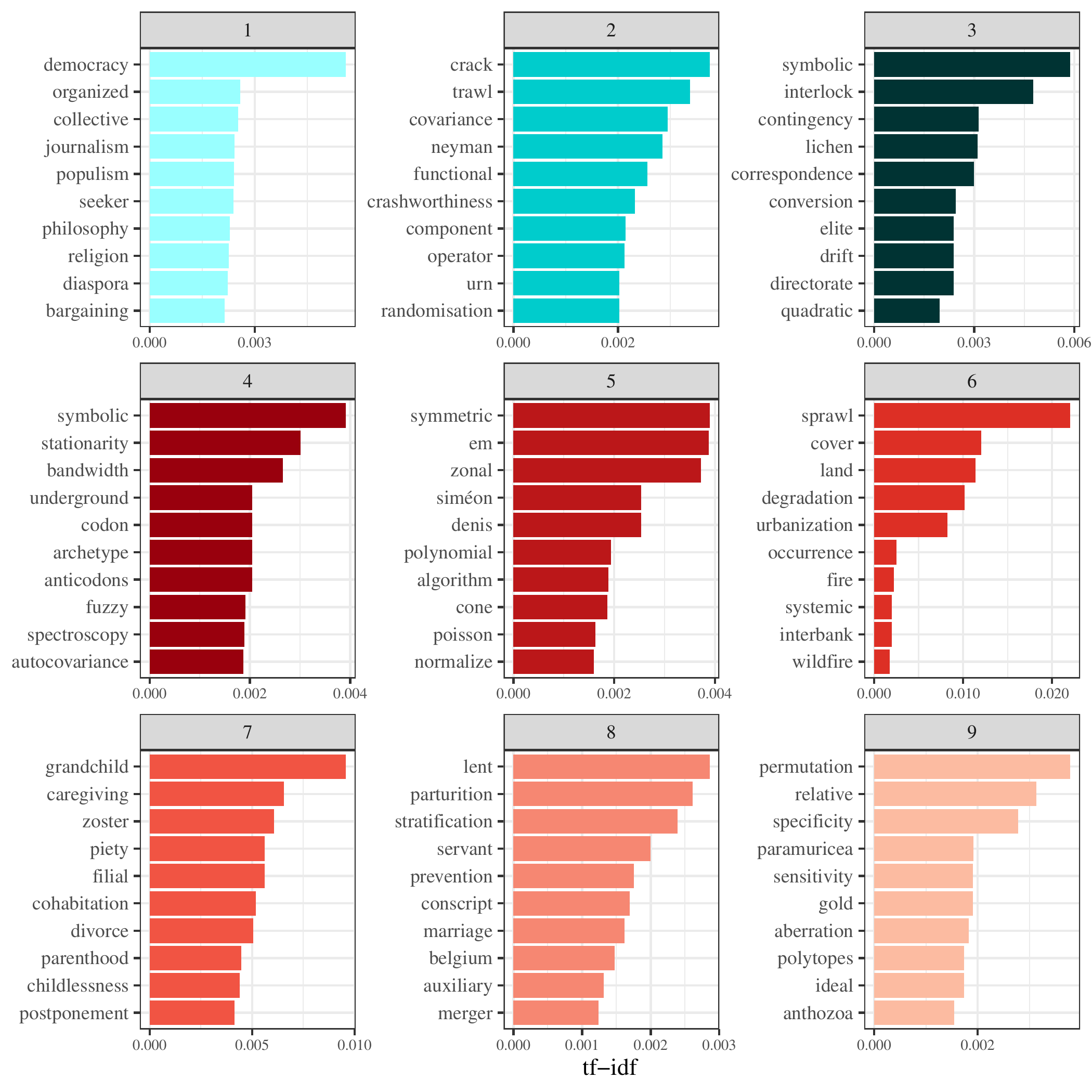}
        \caption{Paper topic}
        \label{fig:sub6}
    \end{subfigure}
    \begin{subfigure}[t]{.495\textwidth}
        \centering
        \includegraphics[width=\textwidth]{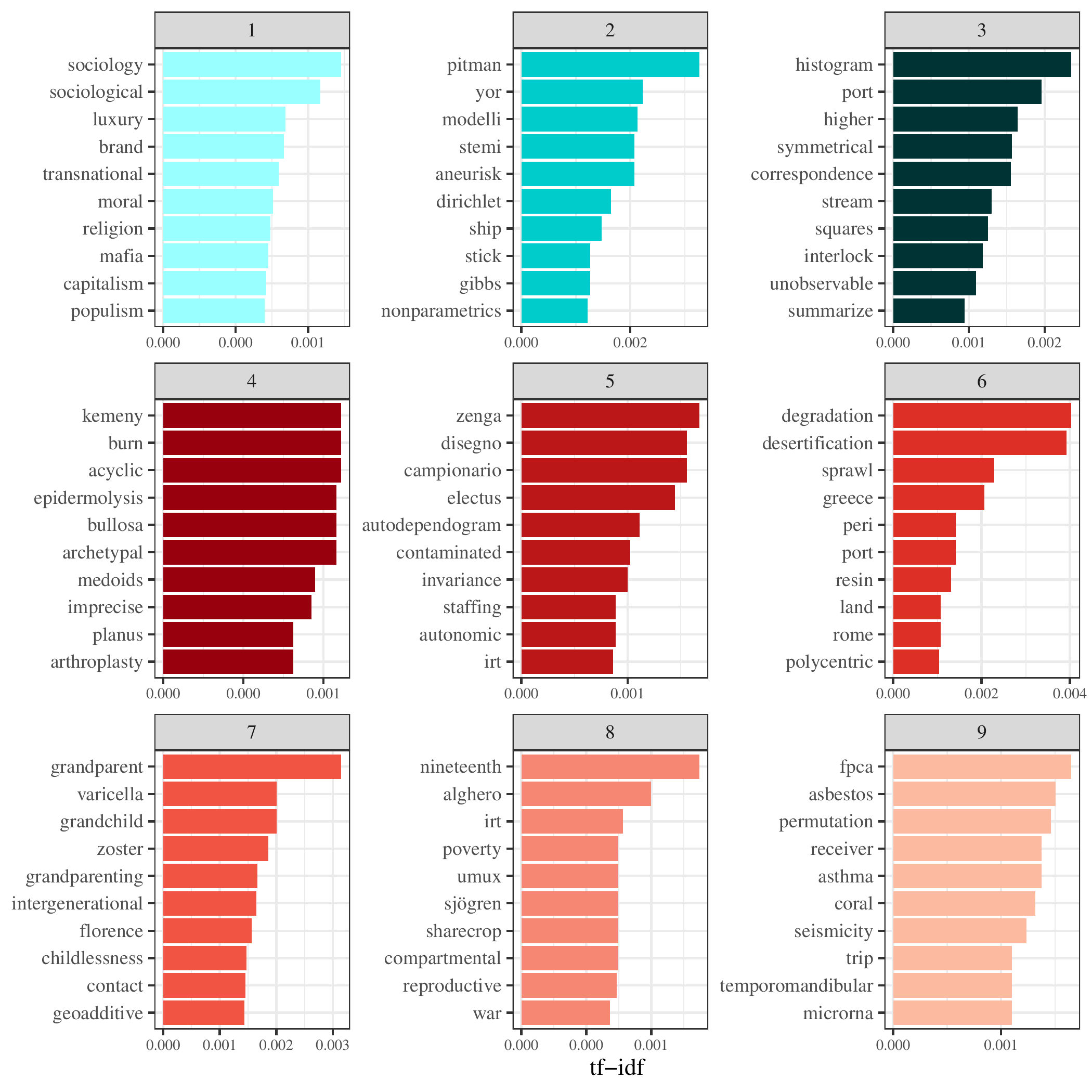}
        \caption{Paper title}
        \label{fig:sub7}
    \end{subfigure}

    \caption{(a) Distribution of topic tf-idf by cluster. (b) Distribution of title tf-idf by cluster.}
    \label{appx:fig4}
\end{figure}

Figure~\ref{appx:fig4} reports the topic tf-idf and title tf-idf distributions for the co-authorship network of Section~\ref{sec:application}; see the main text for discussion.

\end{document}